\documentclass[prodmode,acmtist]{acmsmall} 
\usepackage{graphicx,color,colortbl,psfrag,subfigure}
\usepackage[usenames,dvipsnames]{xcolor}
\usepackage{bbding}
\usepackage{subfigure}
\usepackage{tabularx}
\usepackage{multirow}
\usepackage{array}
\usepackage{flushend}
\usepackage[figuresright]{rotating}
\usepackage{pdflscape}

\usepackage{amssymb}
\usepackage{amsmath}
\usepackage{amsfonts}
\usepackage{mathrsfs}
\usepackage{bigints}

\usepackage[bookmarks=true,
			colorlinks=true,
			linkcolor=black,
			urlcolor=black, 
			citecolor=black]{hyperref}
\usepackage{url}

\usepackage{bm}
\usepackage{algpseudocode}
\usepackage{algorithm}

\acmVolume{7}
\acmNumber{1}
\acmArticle{5}
\acmYear{2015}
\acmMonth{10}

\usepackage[round]{natbib}

\begin{document}

\markboth{B. Chen et al.}{Multi-Keyword Multi-Click Advertisement Option Contracts for Sponsored Search}

\title{Multi-Keyword Multi-Click Advertisement Option Contracts for Sponsored Search}
\author{Bowei Chen
\affil{University College London}
Jun Wang
\affil{University College London}
Ingemar J. Cox
\affil{University of Copenhagen and University College London}
Mohan S. Kankanhalli
\affil{National University of Singapore}}

\begin{abstract} 
In sponsored search, advertisement (abbreviated ad) slots are usually sold by a search engine to an advertiser through an auction mechanism in which advertisers bid on keywords. In theory, auction mechanisms have many desirable economic properties. However, keyword auctions have a number of limitations including: the uncertainty in payment prices for advertisers; the volatility in the search engine's revenue; and the weak loyalty between advertiser and search engine. In this paper we propose a special ad option that alleviates these problems. In our proposal, an advertiser can purchase an option from a search engine in advance by paying an upfront fee, known as the option price. He then has the right, but no obligation, to purchase among the pre-specified set of keywords at the fixed cost-per-clicks (CPCs) for a specified number of clicks in a specified period of time. The proposed option is closely related to a special exotic option in finance that contains multiple underlying assets (multi-keyword) and is also multi-exercisable (multi-click). This novel structure has many benefits: advertisers can have reduced uncertainty in advertising; the search engine can improve the advertisers' loyalty as well as obtain a stable and increased expected revenue over time. Since the proposed ad option can be implemented in conjunction with the existing keyword auctions, the option price and corresponding fixed CPCs must be set such that there is no arbitrage between the two markets. Option pricing methods are discussed and our experimental results validate the development. Compared to keyword auctions, a search engine can have an increased expected revenue by selling an ad option. 
\end{abstract}

\keywords{Sponsored Search, Exotic Option, Pricing Model, Revenue Analysis}

\begin{bottomstuff}

Author's addresses: Bowei Chen and Jun Wang, Department of Computer Science, University College London, Gower Street, London, WC1E 6BT, United Kingdom; Ingemar J. Cox, 1). Department of Computer Science, University of Copenhagen, Sigurdsgade 41, 2200 Copenhagen \O, Denmark 2). Department of Computer Science, University College London, Gower Street, London, WC1E 6BT, United Kingdom; Mohan S. Kankanhalli, Department of Computer Science, School of Computing, National University of Singapore, Singapore 117417, Republic of Singapore. \\
\end{bottomstuff}

\maketitle

\section{Introduction}\label{sec:introduction}

Sponsored search has become an important online advertising format~\cite{iab_2013}, where a search engine sells ad slots in the search engine results pages (SERPs) generated in response to a user's search behaviour. An online user submits a term or phrase within the search box to the search engine. The term or phrase is collectively known as the \emph{query}. The SERP has two types of result listings in response to the submitted query: organic results and paid results. \emph{Organic results} are the Web page listings that most closely match the user\rq{}s search query based on relevance~\cite{Jansen_2011}. \emph{Paid results} are online ads -- the companies who have paid to have their Web pages displayed for certain keywords, so such listings show up when an user submits a search query containing those keywords. The price of an ad slot is usually determined by a keyword auction such as the widely used \emph{generalized second price} (GSP) auction~\cite{Edelman_2007_2,Varian_2007,Lahaie_2007,Borgers_2013,Qin_2014}. In the GSP auction, advertisers bid on keywords present in the query, and the highest bidder pays the price associated with the second highest bid.

Despite the success of keyword auctions, there are two major drawbacks. First, the uncertainty and volatility of bids make it difficult for advertisers to predict their campaign costs and thus complicate their business planning~\cite{Wang_2012_1}. Second, the \lq\lq{}pay-as-you-go\rq\rq{} nature of auction mechanisms does not encourage a stable relationship between advertiser and search engine~\cite{Jank_2010} -- an advertiser can switch from one search engine to another in the next bidding at near-zero cost.

To alleviate these problems, we propose a~\emph{multi-keyword multi-click ad option}. It is essentially a contract between an advertiser and a search engine. It consists of a non-refundable upfront fee, known as the \emph{option price}, paid by the advertiser, in return for the right, but not the obligation, to subsequently purchase a fixed number of clicks for particular keywords for pre-specified fixed cost per clicks (CPCs) during a specified period of time. From the advertiser's perspective, fixing the CPCs significantly reduces the uncertainty in the cost of advertising campaigns. Moreover, for a keyword, if the spot CPC set by keyword auction falls below the fixed CPC of the option contract, the advertiser is not obligated to exercise the option, but can, instead, participate in keyword auctions. Therefore, the option can be considered as an \lq\lq{}insurance\rq\rq{} that establishes an upper limit on the cost of advertising campaigns. From the search engine's perspective, the proposed option is not only an additional service provided for advertisers. We show that the search engine can, in fact, increase the expected revenue in the process of selling an ad option. Also, since the option covers a specific period of time should encourage a more stable relationship between advertiser and search engine. 

An important question for us is to determine the option price and the fixed CPCs associated with candidate keywords in the advertiser's request list. Clearly if the option is priced too low, then significant loss in revenue to the search engine may ensure. Moreover, this may create an arbitrage opportunity where the buyer of the option sells the clicks from their targeted keywords to gain extra profits. Conversely, if the option is priced too high, then the advertiser will not purchase it. In this paper we consider a risk-neutral environment and price the option under the no-arbitrage objective~\cite{Wilmott_2006_1,Bjork_2009}. We use the Monte Carlo method to price the option with multiple candidate keywords and show the closed-form pricing formulas for the cases of single and two keywords. Further, the effects of ad options on the search engine's revenue is analysed. 

This paper has three major contributions. First, we propose a new way to pre-sell ad slots in sponsored search which provides flexible guaranteed deliveries to advertisers. It naturally complements the current keyword auction mechanism and offers both advertiser and search engine an effective risk mitigation tool to deal with fluctuations in the bid price. Although the proposed ad option belongs to a family of exotic options, it differs from existing exotic options that we know from finance and other industries (see Table~\ref{tab:option_comparison} for detailed comparisons): it can be exercised not only once but also multiple times during the contract period; it is not for a single keyword but multiple keywords and each keyword has its own fixed CPC; it allows its buyer to choose which keyword to reserve and advertise at the corresponding fixed CPC later during the contract period. Second, we discuss a generalized pricing method for the proposed ad option (see Algorithm~\ref{algo:solution}) to deal with the high dimensionality. Third, we demonstrate that, compared to keyword auctions, a search engine can have an increased expected revenue by selling an ad option.

The rest of the paper is organised as follows. Section~\ref{sec:related_work} reviews the related literature. Section~\ref{sec:option_contract} introduces the design of proposed ad option, discusses the option pricing methods and analyses the option effects on the search engine's revenue. Section~\ref{sec:experiment} presents our empirical evaluation and Section~\ref{sec:conclusion} concludes the paper. Several important mathematical results are provided in Appendices~\ref{app:proof_no_early_exercise}-\ref{app:special_cases}.
 
\section{Related Work}\label{sec:related_work}

The work presented in this paper touches upon several streams of literature. We first review the prior work on options in finance and other industries, and then discuss the related literature in guaranteed advertising deliveries.

\subsection{Options and Their Pricing Methods} 

Options have been known and traded for many centuries and can be traced back to the 17th century~\cite{Constantinides_2001}. 
A \emph{standard option} is a contract in which the seller grants the buyer the right, but not the obligation, to enter into a transaction with the seller to either buy or sell an underlying asset at a fixed price on or prior to a fixed date. The fixed price is called the \emph{strike price} and the fixed date is called the \emph{expiration date}. The seller grants this right in exchange for a certain amount of money, called the \emph{option price}. An option is called the \emph{call option} or \emph{put option} depending on whether the buyer is purchasing the right to buy or sell the underlying asset. The simplest option is the \emph{European option}~\cite{Wilmott_2006_1}, which can be exercised only on the expiration date. This differs from the \emph{American option}~\cite{Wilmott_2006_1}, which can be exercised at any time during the contract lifetime. Both European and American options are called~\emph{standard options}. 


In the beginning of the 1980s, standard options became more widely understood and their trading volume increased dramatically. Financial institutions began to search for alternative forms of options, known as~\emph{exotic options}~\citep{Zhang_1998}, to meet their new business needs. Among them, two types of options, multi-asset options and multi-exercise options, are particularly relevant to our research.

\begin{sidewaystable}[htp]
\tbl{Comparison of the proposed ad option and other options. The price of the $i$th underlying asset/keyword at time $t$ is denoted by $C_i(t)$, where $t$ is a continuous time point in period $[0, T]$ and $T$ is the contract expiration date; if there is only one underlying asset we denote its price by $C(t)$. The strike/fixed payment price, of the $i$th underlying asset/keyword is denoted by $F_i$; if there is only one strike price we denote it by $F$. The weight of $i$th asset/keyword in a basket-type option is denoted by $\omega_i$. Note that in the $n$-keyword $1$-click ad option, $\omega_{j i}$ represents the weight of the $i$th broad matched keyword for the $j$th candidate keyword, and $k_j$ represents the number of broad matched keywords. Detailed descriptions of notations are provided in Table~\ref{tab:notations}.\label{tab:option_comparison}}{
\begin{tabular}{|c|c|c|c|c|c|c|}
\hline
\multirow{2}{*}{Option contract}  
& 
\multirow{2}{*}{Payoff function} & \hspace{-5pt} Underlying  \hspace{-5pt} & Exercise  & Early  & Strike & Application\\
	      &   &  variable & \hspace{-5pt} opportunity \hspace{-5pt} & \hspace{-5pt} exercise \hspace{-5pt} & price  & area\\
\hline
\rowcolor{Gray!50} 
$n$-keyword $1$-click ad option
& $\max \{C_1(t) - F_1, \ldots, C_n(t) - F_n, 0\}$ 
& Multiple
& Single 
& Yes 
& \hspace{-5pt} Multiple \hspace{-5pt}
& Keywords\\
\rowcolor{Gray!50} 
(keyword exact or broad match) & & & & & & \\
\hline
\rowcolor{Gray!50} 
$n$-keyword $1$-click ad option
& 
\hspace{-5pt}
\parbox{2.75in}{
$$
\max \bigg\{\sum_{i = 1}^{k_1} \omega_{1i} C_{1i}(t) - F_1, \cdots, \sum_{i = 1}^{k_n} \omega_{ni} C_{ni}(t) - F_n, 0\bigg\}
$$}
\hspace{-5pt}
& Multiple
& Single 
& Yes 
& \hspace{-5pt} Multiple \hspace{-5pt}
& Keywords\\
\rowcolor{Gray!50} 
(keyword broad match) & & & & & & \\
\hline
European standard call option
& \multirow{2}{*}{$\max \{C(T) - F, 0\}$}
& \multirow{2}{*}{Single} 
& \multirow{2}{*}{Single} 
& \multirow{2}{*}{No} 
& \multirow{2}{*}{Single}
& Equity stock, \\
\cite{Wilmott_2006_1} & & & & & & or index\\
\hline
American standard call option
& \multirow{2}{*}{$\max \{C(t) - F, 0\}$}
& \multirow{2}{*}{Single} 
& \multirow{2}{*}{Single} 
& \multirow{2}{*}{Yes} 
& \multirow{2}{*}{Single}
& Equity stock,\\
\cite{Wilmott_2006_1} & & & & & & or index\\
\hline
& \multirow{5}{*}{\parbox{1.5in}{$$\max \bigg\{\sum_{i = 1}^n \omega_i C_i(T) - F, 0 \bigg\}$$}}
& \multirow{5}{*}{Multiple} 
& \multirow{5}{*}{Single} 
& \multirow{5}{*}{No}
& \multirow{5}{*}{Single} 
& Index of \\ 
European & & & & & & equity stocks \\
basket call option & & & & & & bonds or\\
\cite{Krekel_2006} & & & & & & foreign \\
 & & & & & & currencies\\
\hline
European  
& \multirow{3}{*}{$\max \{C_1(T) - F_1, C_2(T) - F_2, 0\}$}
& \multirow{3}{*}{Double} 
& \multirow{3}{*}{Single} 
& \multirow{3}{*}{No}
& \multirow{3}{*}{Double}
& \\
dual-strike call option & & & & & & \\
\cite{Zhang_1998} & & & & & &  Equity stocks,\\
\cline{1-6}
European 
& \multirow{3}{*}{$\max \{\max\{ C_1(T) , \ldots, C_n(T) \} - F, 0\}$}
& \multirow{3}{*}{Multiple} 
& \multirow{3}{*}{Single} 
& \multirow{3}{*}{No}
& \multirow{3}{*}{Single}
& or indexes of \\
rainbow call on max option & & & & & & equity stocks,\\
\cite{Ouwehand_2006} & & & & & & or bonds, or\\
\cline{1-6}
European 
& \multirow{3}{*}{$\max \{C_1(T), C_2(T), F\}$}
& \multirow{3}{*}{Double} 
& \multirow{3}{*}{Single} 
& \multirow{3}{*}{No}
& \multirow{3}{*}{Single}
& foreign\\ 
paying the best and cash option & & & & & & currencies\\
\cite{Johnson_1987} & & & & & & \\
\cline{1-6}
European quotient call option
& \multirow{2}{*}{$\max \{C_1(T)/C_2(T) - F, 0\}$}
& \multirow{2}{*}{Double} 
& \multirow{2}{*}{Single} 
& \multirow{2}{*}{No}
& \multirow{2}{*}{Single}
&  \\ 
\cite{Zhang_1998} & & & & & & \\
\hline
\end{tabular}
}
\end{sidewaystable}

\emph{Multi-asset options} are the options written on at least two underlying assets~\citep{Zhang_1998}. These underlying assets can be stocks, bonds, currencies and indices in either the same category or different markets. Several types of multi-asset options are worth mentioning, such as basket options, dual-strike options, rainbow options, paying the best and cash options, and quotient options. Table~\ref{tab:option_comparison} provides a brief summary of these multi-asset options, and compares them to standard options and our proposed multi-keyword multi-click ad options (see Section~\ref{sec:option_contract}) along the following seven dimensions: payoff function, underlying variable, exercise opportunity, early exercise opportunity, strike price and application area. The comparison indicates that our proposed ad options is more complex than previous proposals.

In Table~\ref{tab:option_comparison}, it is worth emphasising basket options and dual-strike options. \emph{Basket options} are those options whose payoff is determined by the weighted sum of underlying asset prices~\citep{Wilmott_2006_1}. This structure can be extended to the keyword broad match setting\footnote{The keyword match type setting helps the search engine to control which searches can trigger an advertiser's ad. Under the \emph{exact match setting}, the advertiser's ad may show on searches that are an exact term and close variations of that exact term; Under the \emph{broad match setting}, the advertiser's ad may show on searches that include misspellings, synonyms, other relevant variations and related searches. For further details, see \href{https://support.google.com/adwords/}{https://support.google.com/adwords/}}, where the weights are the probabilities that sub-phrases occur in search queries. \emph{Dual-strike options} are options with two different strike prices for two different underlying assets~\citep{Zhang_1998}. One simple version of our proposed ad options is a dual-strike call option, which allows an advertiser to switch between his targeted two keywords during the contract lifetime. However, in sponsored search, the number of candidate keywords to choose from is usually more than two, so the two keywords are extended to higher dimensions. In addition, as an advertiser usually needs more than a single click for guaranteed delivery, the dual-strike call option is extended to a multi-exercise option.  

\emph{Multi-exercise options} are a generalisation of American options, which provide a buyer with more than one exercise right and sometimes control over one or more other variables~\citep{Villinski_2004}, e.g., the amount of the underlying asset exercised in certain time periods. Multi-exercise options have become more prevalent over the past decade, particularly, in the energy industry, such as electricity swing options and water options. Contributors to the multi-exercise options include~\citet{Deng_2000}, \citet{Deng_2006}, \citet{Clewlow_2000}, \citet{Villinski_2004}, \citet{Weron_2006}, \citet{Marshall_2011} and~\citet{Marshall_2012}. Their work is not further discussed here as our proposed ad option is a simple example of multi-exercise options. Compared to the energy industry, the multi-exercise opportunity in sponsored search is more flexible. Advertisers are allowed to exercise options at any time in the option lifetime, i.e. the exercise time is not pre-specified, and no minimum number of clicks is required for each exercise. Therefore, there is no penalty fee if the advertiser does not exercise the minimum clicks. In addition, there is no transaction fee for ad options in sponsored search. 

Motivated by an attempt to model the fluctuations of asset prices, Brownian motion (i.e., the continuous-time random walk process~\cite{Shreve_2004_1}) was first introduced by~\citet{Bachelier_1900} to price an option. However, the impact of his work was not recognised by financial community for many years. Sixty five years later, \citet{Samuelson_1965} replaced Bachelier's assumptions on asset price with a geometric form, called the \emph{geometric Brownian motion (GBM)}. In the GBM model, the proportional price changes are exponentially generated by a Brownian motion. While the GBM model is not appropriate for all financial assets in all market conditions, it remains the reference model against which any alternative dynamics are judged.

The research of Samuelson highly affected \citet{Black_1973} and~\citet{Merton_1973}, who then examined the option pricing based on a GBM. They constructed a portfolio from risky and risk-less underlying assets to replicate the value of an European option. Risky assets can be stocks, foreign currencies, indices, and so on; risk-less assets can be bonds. Once the risky part of the replicated portfolio is estimated, the option value can be obtained accordingly. The pricing methods proposed by \citet{Black_1973} and~\citet{Merton_1973} were based on the assumption that investors on the market cannot obtain arbitrage. Therefore, the replicated portfolio is treated as a self-adjusting process whose least expectation of returns increase at the same speed as the constant bank interest rate. If considering the constant bank interest rate as a discount factor, the discounted value of the replicated portfolio would be a martingale~\citep{Bjork_2009}, whose probability measure is called the~\emph{risk-neutral probability measure}. Since a closed-form pricing formula can be obtained from the settings of \citet{Black_1973} and~\citet{Merton_1973}, we normally call their work as the \emph{Black-Scholes-Merton (BSM) option pricing formula}. The BSM option pricing formula spurred research in this field. Various numerical procedures then appeared, including lattice methods, finite difference methods and Monte Carlo methods. These numerical procedures are capable of evaluating more complex options when the closed-form solution does not exist. In this paper, the Monte Carlo method we discussed can quickly price an ad option where the number of candidate keywords is larger than two.

\subsection{Guaranteed Advertising Deliveries} 

Guaranteed contracts appeared in the early stages of online advertising (particularly in display advertising). They were mostly negotiated by advertisers and publishers\footnote{Publishers are sellers in display advertising.} privately~\cite{Edelman_2007_2}. Each negotiation contains an amount of needed display impressions over a certain period of time and a pre-specified guaranteed price. Hence, in discussing guaranteed deliveries, the following issues must be considered: allocation and pricing. Many studies discussed these two issues separately. Allocation models is reviewed first, then the pricing models.

\citet{Feldman_2009} studied an ad selection algorithm for a publisher whose objective is not only to fulfil the guaranteed contracts but also to deliver the well-targeted display impressions to advertisers. This research was more relevant to a service matching problem. The allocation of impressions between the guaranteed and non-guaranteed channels was first discussed by~\citet{Ghosh_2009}, where a publisher was considered to act as a bidder who bids for guaranteed contracts. This modelling setting was reasonably good as the publisher acts as a bidder who would allocate impressions to online auctions only when other winning bids are high enough. \citet{Balseiro_2011} investigated the same allocation problem but used some stochastic control models. In their model, for a given price of an impression, the publisher can decide whether to send it to ad exchanges or assign it to an advertiser with a fixed reserve price. The decision making process aims to maximise the expected total revenue. \citet{Roels_2009} proposed a similar allocation framework to~\citet{Balseiro_2011}, where the publisher can dynamically select which guaranteed buy requests to accept and to deliver the guaranteed impressions accordingly. However, compared to~\citet{Balseiro_2011}, the uncertainty in advertisers\rq{} buy requests and the traffic of a website were explicitly modelled under the revenue maximisation objective. Recently, a lightweight allocation framework was proposed by~\citet{Bharadwaj_2012}. They used a simple greedy algorithm to simplify the computations of revenue maximisation.

Two algorithms for pricing the guaranteed display contracts were discussed by~\citet{Bharadwaj_2010}. Each contract has a large number of impressions and the proposed algorithms solved the revenue optimisation problem for the given number of user visits (i.e., the demand level). However, their work did not consider the auction effects on the contract pricing, and the developed algorithms were purely based on the statistics of users' visits. 

Consider the case where the online advertising market is bouyant (i.e., the winning payment prices for specific ad slots from online auctions increase) and non-guaranteed selling becomes more profitable for publishers. In this case, they may want to cancel the sold guaranteed contracts before the time that the targeted impressions will be created. Online auctions with cancellations were recently discussed by~\citet{Babaioff_2009} and \citet{Constantin_2009}. They both considered a design where a publisher can cancel the sold guaranteed contracts but needs to pay a penalty to the advertisers. The proposed auctions with cancellations exhibit interesting economic properties, such as allocative efficiency and equilibrium solution. However, there may exist speculators who pursue the cancellation penalty only. In fact, the discussed cancellation penalty is very similar to over-selling of flight tickets~\cite{Talluri_2004}. 

\citet{Salomatin_2012} studied a framework of guaranteed deliveries for sponsored search, under which advertisers are able to send their guaranteed requests to a search engine. Each guaranteed request includes the needed number of clicks and the ad budget. The search engine then decides guaranteed deliveries according to search queries and available positions. Since the allocation decision is based on the joint revenue maximisation from guaranteed deliveries and keyword auctions, some advertisers may not receive all their demanded clicks. In such cases, the search engine pays a penalty. However, advertisers still have less control of the ad exposure time and the position of the ad. In addition, with the number of guaranteed advertisers increasing, it is less likely that advertisers can meet their business needs with such a mechanism.

The concept of ad option was initially introduced by~\citet{Moon_2010} (even though \citet{Meinl_2009}~discussed the possibility of Web service derivatives, their proposal was not intended for online advertising). \citet{Moon_2010} proposed that the ad option buyer can be guaranteed the right to choose the minimum payment between cost-per-mille (CPM) and CPC once click-through rate (CTR) is realized. This option structure was similar to a \emph{paying the worst and cash option}~\cite{Zhang_1998}. In addition, \citet{Moon_2010}~suggested option pricing under the framework of a Nash bargaining game. Simply, they considered two utility functions: one for the advertiser and one for the publisher. The objective function is the product of these two utilities and each utility function is restricted by a negotiation power. Therefore, the option price is the optimal solution which maximises the negotiated join utility. Another ad option was discussed by~\citet{Wang_2012_1} (and later \citet{Chen_2014_3}) for display advertising. The option allows its buyer to select his preferred payment scheme (either CPM or CPC) for the fixed payment. For example, an advertiser can choose to pay a fixed CPC for targeted display impressions.  They discussed the lattice methods for option pricing and investigated the stochastic volatility (SV) model for the cases where the GBM assumption is not valid empirically. However, their work was limited to an univariate case as the SV model cannot be easily extended to multiple variables based on the lattice framework.
\section{Multi-Keyword Multi-Click Ad Options}
\label{sec:option_contract}

We first introduce how a multi-keyword multi-click ad option works, then discuss the option pricing methods, and finally provide an analysis of the search engine's revenue.

\subsection{Guaranteed Delivery in Sponsored Search via Ad Options}
\label{sec:option_structure}

\begin{figure}[tp]
\centering
\includegraphics[width=1\linewidth]{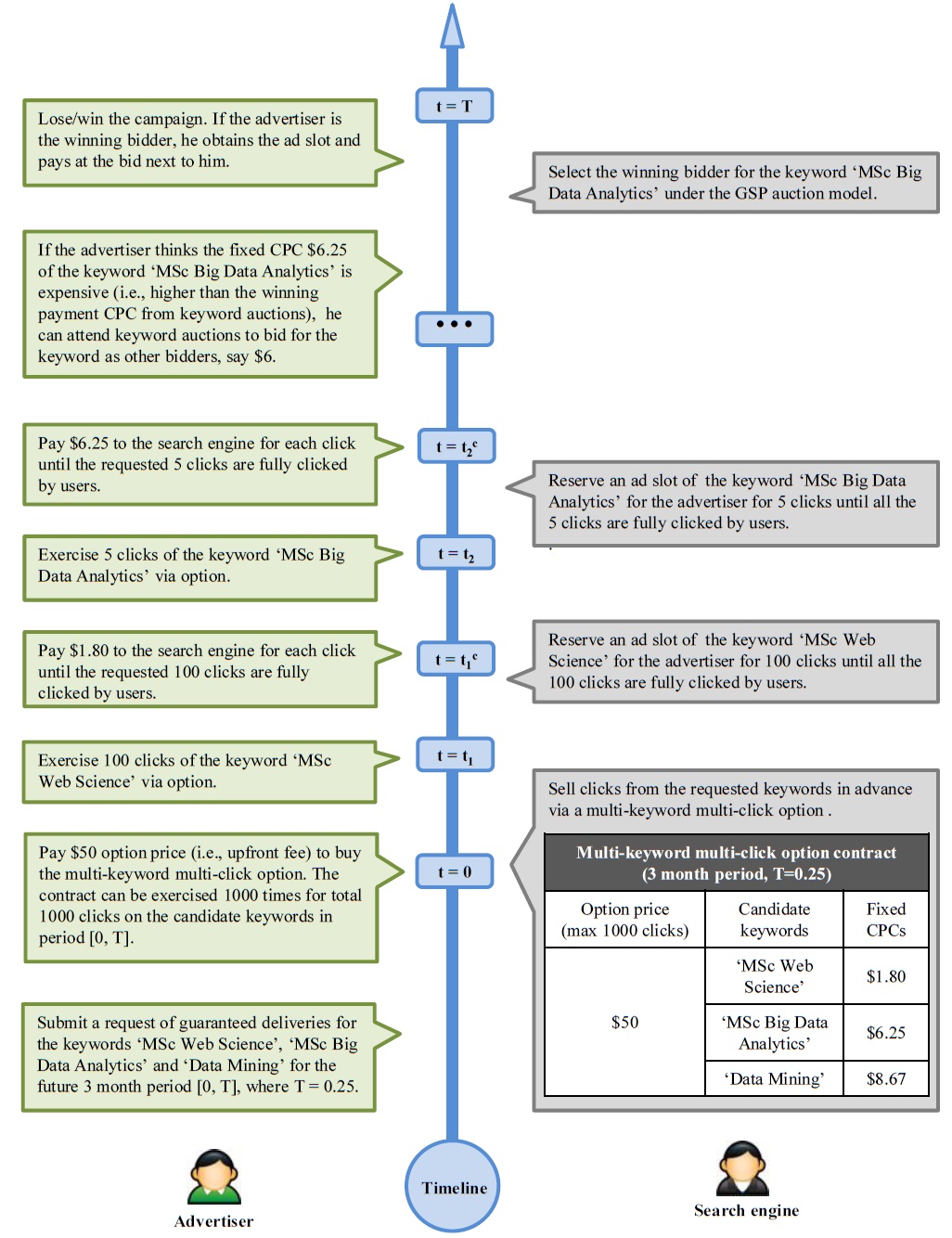}
\caption{Schematic view of buying, selling and exercising a multi-keyword multi-click ad option in sponsored search.}
\label{fig:option_structure}
\end{figure}

We use the following example to illustrate our idea. Suppose that a computer science department creates a new master degree programme on \lq{}Web Science and Big Data Analytics\rq{} and is interested in an advertising campaign based around relevant search terms such as \lq{}MSc Web Science\rq{}, \lq{}MSc Big Data Analytics\rq{} and \lq{}Data Mining\rq{}, etc. The campaign is to start immediately and last for three months and the goal is to generate at least 1000 clicks on the ad which directs users to the homepage of this new master programme. The department (i.e., advertiser) does not know how the clicks will be distributed among the candidate keywords, nor how much the campaign will cost if based on keyword auctions. However, with the ad option, the advertiser can submit a request to the search engine to lock-in the advertising cost. The request consists of the candidate keywords, the overall number of clicks needed, and the duration of the contract. The search engine responds with a price table for the option, as shown in Figure~\ref{fig:option_structure}. It contains the option price and the fixed CPC for each keyword. The CPCs are fixed yet different across the candidate keywords. The contract is entered into when the advertiser pays the option price.

During the contract period $[0, T]$, where $T$ represents the contract expiration date (and is three months in this example), the advertiser has the right, at any time, to exercise portions of the contract, for example, to buy a requested number of clicks for a specific keyword. This right expires after time $T$ or when the total number of clicks have been purchased, whichever is sooner.  For example, at time $t_1 \leq T$ the advertiser may exercise the right for 100 clicks on the keyword \lq{}MSc Web Science\rq{}. After receiving the exercise request, the search engine immediately reserves an ad slot for the keyword for the advertiser until the ad is clicked on 100 times. In our current design, the search engine decides which rank position the ad should be displayed as long as the required number of clicks is fulfilled - we assume there are adequate search impressions within the period. It is also possible to generalise the study in this paper and define a rank specific option where all the parameters (CPCs, option prices etc.) become rank specific. The advertiser can switch among the candidate keywords and also monitor the keyword auction market. If, for example, the CPC for the keyword \lq{}MSc Web Science\rq{} drops below the fixed CPC, then the advertiser may choose to participate in the auction rather than exercise the option for the keyword. If later in the campaign, the spot price for the keyword \lq{}MSc Web Science\lq{}' exceeds the fixed CPC, the advertiser can then exercise the option.

Figure~\ref{fig:option_structure} illustrates the flexibility of the proposed ad option. Specifically, (i) the advertiser does not have to use the option and can participate in keyword auctions as well, (ii) the advertiser can exercise the option at any time during the contract period, (iii) the advertiser can exercise the option up to the maximum number of clicks, (iv) the advertiser can request any number of clicks in each exercise provided the accumulated number of exercised clicks does not exceed the maximum number, and (v) the advertiser can switch among keywords at each exercise at no additional cost. Of course, this flexibility complicates the pricing of the option, which is discussed next.

\subsection{Option Pricing Methods}\label{sec:option_pricing}

The proposed multi-keyword multi-click ad option enables an advertiser to fix his advertising cost and construct a set of candidate keywords beforehand, yet leave the decision of selecting suitable keywords for matching and the exact timing to place the ad to later. Since the advertiser enjoys great flexibility in sponsored search, there is an intrinsic value associated with an ad option and the buyer needs to pay an upfront option price first. In the following discussion, we focus on calculating a fair upfront option price for the given option candidate keywords, the current winning payment prices, the volatility of these keywords, the length of contract period, the risk-less bank interest rate, and the fixed CPCs for candidate keywords. Note that the fixed CPCs are considered as given variables as they can be set by the search engine after receiving the advertiser's request or be proposed by the advertiser in his request. Either case will not affect our valuation of the option. We follow the scenario of the motivating example presented in Figure~\ref{fig:option_structure} and consider the search engine sets the fixed CPCs. 

Recall that Table~\ref{tab:option_comparison} presents two different payoff functions for the proposed ad option. The first payoff function can be used to price an ad option with either the keyword exact or broad match setting, which is determined by what is the match type of the winning payment prices used. However, if only having the exact match winning payment prices from keyword auctions and the advertiser wants to have an ad option with keyword broad match setting, the second payoff function can be used for option pricing. In the following, we discuss the option pricing based on the first payoff function. Same method can be applied to the second payoff function, for further details see Section~\ref{sec:discussion}. 

\subsubsection{Underlying Stochastic Model}

The winning payment CPC of the candidate keyword $K_i$ (for a specific slot/position) at time $t$ is denoted by $C_i(t)$. Its movement can be described by a multivariate geometric Brownian motion (GBM)~\cite{Samuelson_1965}:  
\begin{align}
d C_i (t) = & \ \mu_i C_i (t) dt + \sigma_i C_i (t) d W_i (t), \ \ \ i=1,\ldots, n, \label{eq:sde_gbm_multi}
\end{align}
where $\mu_i$ and $\sigma_i$ are constants representing the drift and volatility of the CPC respectively, and $W_i (t)$ is a standard Brownian motion satisfying the conditions:
\begin{align}
\mathbb{E}(d W_i (t)) = & \ 0, \nonumber \\
\mathrm{var}(d W_i (t)) = & \ \mathbb{E}(d W_i (t) d W_i (t)) = dt,  \nonumber \\
\mathrm{cov}(d W_i (t), d W_j (t)) = & \ \mathbb{E}(d W_i (t) d W_j (t)) = \rho_{i j} dt,  \nonumber 
\end{align}
where $\rho_{i j}$ is the correlation coefficient between keywords $K_i$ and $K_j$, such that $\rho_{ii} = 1$ and $\rho_{i j} = \rho_{j i}$. The correlation matrix is denoted by $\boldsymbol\Sigma$, so that the covariance matrix is simply $\boldsymbol M \boldsymbol\Sigma \boldsymbol M$, where $\boldsymbol M$ is the matrix with the $\sigma_i$ along the diagonal and zeros everywhere else. For the reader's convenience, detailed descriptions of notations are provided in Table~\ref{tab:notations}.

\begin{table}[t]
\tbl{Summary of notations.\label{tab:notations}}{
\begin{tabular}{|l|p{3.75in}|}
\hline
\multicolumn{1}{|c|}{Notation} & \multicolumn{1}{c|}{Description}\\
\hline
$r$ & Constant continuous (risk-less) interest rate.\\
$T$ & Option expiration date.\\
$t$ & Continuous time point in $[0,T]$.\\
$m$ & Number of total clicks specified by an ad option.\\
$n$ & Number of total number of keywords specified by an ad option.\\
$\boldsymbol{K}$ & Keywords specified by an ad option, $\boldsymbol{K} = \{K_1, \ldots, K_n\}$.\\
$\boldsymbol{F}$ & Pre-specified fixed CPCs for keywords $\boldsymbol{K}$.\\
$\boldsymbol{C}(t)$ & Winning payment CPCs for keywords $\boldsymbol{K}$ from auctions at time $t$.\\
$V(t, \boldsymbol{C}(t); T, \boldsymbol{F}, m)$ & Value of an $n$-keyword $m$-click ad option at time $t$.\\
$\mu_i$ & Constant drift of CPC for keyword $K_i, i = 1,\ldots, n$.\\
$\sigma_i$ & Constant volatility of CPC for keyword $K_i, i = 1,\ldots, n$.\\
$W(t)$ & Standard Brownian motion at time $t$.\\
$\boldsymbol\Sigma$ & Price correlation matrix, in which $\rho_{i j}$ is the correlation coefficient between keywords $K_i$ and $K_j$, such that $\rho_{ii} = 1$ and $\rho_{i j} = \rho_{j i}$.\\
$\boldsymbol M \boldsymbol\Sigma \boldsymbol M$ & Price covariance matrix, where $\boldsymbol M$ is the matrix with $\sigma_i$ along the diagonal and zeros everywhere else. \\
$\Phi(\boldsymbol{C}(t))$ & Payoff function of an ad option at time $t$.\\
$\pi_0$ & Option price (i.e., upfront fee) of an ad option.\\
$N(\mu, \sigma^2)$ & Normal distribution with mean $\mu$ and variance $\sigma^2$.\\
$MVN(\boldsymbol\mu, \boldsymbol M \boldsymbol\Sigma \boldsymbol M)$ &
Multivariate normal distribution with mean $\boldsymbol\mu$ and variance $\boldsymbol M \boldsymbol\Sigma \boldsymbol M$.\\
$\mathscr{N}[\cdot]$ & Cumulative probability distribution of a standard normal distribution.\\
\hline
\end{tabular}
}
\end{table}

Since the GBM assumption lays the foundation of pricing the proposed ad option, we provide several discussions and investigations of it. In Section~\ref{sec:discussion}, we explain why the GBM assumption is suitable for pricing an ad option in sponsored search, and also highlight its limitations.  In Section~\ref{sec:exp_para_estimation}, we discuss the estimation of GBM parameters. In Section~\ref{sec:model_validation}, we conduct goodness-of-fit tests with real datasets and track the \lq\lq{}errors\rq\rq{} of the calculated option price when the GBM assumption is not valid empirically.

\subsubsection{Terminal Value Pricing}

To simplify the discussion and without loss of generality,  the value of an $n$-keyword $m$-click ad option can be decomposed as the sum of $m$ independent $n$-keyword $1$-click ad options. If an advertiser buys an ad option at time $0$, the option price $\pi_0$ can be expressed as follows
\begin{equation}
\pi_0 = V(0, \boldsymbol{C}(0); T, \boldsymbol{F}, m) = m V(0, \boldsymbol{C}(0); T, \boldsymbol{F}, 1), \label{eq:option_price_1}
\end{equation}
where $V(0, \boldsymbol{C}(0); T, \boldsymbol{F}, m)$ represents the option value at time $0$. 

Our focus now centres on the $n$-keyword $1$-click ad option. Adopting the basic economic setting~\cite{Narahari_2009}, we assume that an advertiser is risk-neutral. In other words, he has no preference across the candidate keywords and exercises the option for the keyword which has the maximum difference between its winning payment price and the pre-specified fixed price. This difference shows the value of the option because the advertiser is offered the right to move from the auction market to the guaranteed market. 

Let us first consider if the advertiser exercises the option at the contract expiration date $T$, the option payoff can be defined as follows
\begin{align}
 \Phi(\boldsymbol{C}(T))  =  & \ \max\{C_1(T)-F_1, \ldots, C_n(T) - F_n, 0\}.
 \label{eq:option_payoff_T}
\end{align}
Note that the option payoff in sponsored search does not mean the direct reward but it measures the difference of advertising cost between the auction market and the guaranteed market. By having Eq.~(\ref{eq:option_payoff_T}), we can see if the advertiser would like to exercise the option early by using the backward deduction method. The option value at time  time $t < T$ is then
\begin{equation*}
 V(t, \boldsymbol{C}(t); T, \boldsymbol{F},  1)  = 
 \Bigg\{
 \begin{array}{ll}
 \Phi(\boldsymbol{C}(t))  , & \textrm{if early exercise}, \\
 \mathbb{E}^{\mathbb{Q}}_t \big[ e^{-  r (T-t)} \Phi(\boldsymbol{C}(T))  \big], & \textrm{if not early exercise}, \\
\end{array}
\end{equation*}
where $r$ is the constant risk-less bank interest rate and $\mathbb{E}^{\mathbb{Q}}_t[\cdot]$ is the conditional expectation with respect to time $t$ under the probability measure $\mathbb{Q}$. As we use the risk-less bank interest rate as the discounted factor, the probability measure $\mathbb{Q}$ is also called the \emph{risk-neutral probability measure}~\cite{Bjork_2009}. Appendix~\ref{app:derivation_bsm} discusses the rationale for using the risk-less bank interest rate and introduces an alternative method of option pricing.

Let us now return to the decision making problem. If the ad option is exercised early at time $t$, the option value is equal to its payoff $\Phi(\mathbf{C}(t))$. However, if the ad option is not exercised, the option value at time $t$ is equal to the discounted value of the expected payoff at the expiration date $T$. The comparison between $\Phi(\boldsymbol{C}(t))$ and $\mathbb{E}^{\mathbb{Q}}_t \big[ e^{-  r (T-t)} \Phi(\boldsymbol{C}(T)) \big]$ informs the optimal decision for the advertiser. Since the payoff function defined is convex, we then obtain the following inequality (see Appendix~\ref{app:proof_no_early_exercise}): 
\begin{equation}
\Phi(\boldsymbol{C}(t)) \leq \mathbb{E}^{\mathbb{Q}}_t \big[ e^{-  r (T-t)} \Phi(\boldsymbol{C}(T)) \big].
\label{eq:no_early_exercise}
\end{equation}
Eq.~(\ref{eq:no_early_exercise}) illustrates, to gain the maximum option value, the advertiser will not exercise the option until its expiration date. Hence, the option price should be computed at the discounted value of the expected payoff from the expiration date $T$. Together with Eq.~(\ref{eq:option_price_1}), we can obtain the option pricing formula for the $n$-keyword $m$-click ad option:
\begin{align}\label{eq:option_price}
 \pi_0  = 
 & \ m e^{-  r T} \mathbb{E}^{\mathbb{Q}}_0 
 \big[ \Phi(\boldsymbol{C}(T)) \big].
\end{align}

It is worth noting that we rule out arbitrage~\cite{Varian_1987} between the auction market and the guaranteed market in option pricing. The concept of arbitrage can be understood as the \lq\lq{}free lunch\rq\rq{}. As a market designer, we need to make sure that everyone obtains something by paying something so that it is fair to both the buy and sell sides. Since we assume that an advertiser is risk-neutral, the risk-less bank interest rate can be employed as the benchmark rate to rule out arbitrage. Eq.~(\ref{eq:option_price}) can also be obtained by constructing an advertising strategy for the advertiser as discussed in Appendix~\ref{app:derivation_bsm}. 

\subsubsection{Solutions}

Eq.~(\ref{eq:option_price}) can be expanded in integral form as follows
\begin{align}
\pi_0  =  
  & \ m e^{-r T} \big(2 \pi T \big)^{-\frac{n}{2}} |\boldsymbol\Sigma |^{-\frac{1}{2}} \Bigg( \prod_{i=1}^n \sigma_i \Bigg)^{-1} \nonumber \\
  & \ \ \ \ \times
  \bigintsss_{0}^{\infty} \cdots \bigintsss_{0}^{\infty} \frac{\Phi(\boldsymbol{\widetilde{C}})}{\prod_{i=1}^n \widetilde{C_i} } \exp\left\{ -\frac{1}{2} \boldsymbol\zeta^T{\boldsymbol\Sigma}^{-1} \boldsymbol\zeta\right\} d \boldsymbol{\widetilde{C}}, \label{eq:option_price_integral}
\end{align}
where $\boldsymbol\zeta = (\zeta_1, \ldots, \zeta_n)^\prime$, $\zeta_i = \frac{1}{\sigma_i \sqrt{T}} 
\big( \ln \{\widetilde C_i/C_i(0)\} - (r-\frac{\sigma_i^2}{2}) T \big)$, and other notations are described in Table~\ref{tab:notations}.

Closed form solutions to Eq.~(\ref{eq:option_price_integral}) can be derived if $n \leq 2$. If $n=1$, Eq.~(\ref{eq:option_price_integral}) is equivalent to the Black-Scholes-Merton (BSM) pricing formula for an European call option~\cite{Black_1973,Merton_1973}. If $n=2$, Eq.~(\ref{eq:option_price_integral}) contains a bivariate normal distribution and the option price can be obtained by employing the pricing formula for a dual-strike European call option~\cite{Zhang_1998}. The closed form solutions are provided in Appendix~\ref{app:special_cases}.
 
For $n \geq 3$, taking integrals in Eq.~(\ref{eq:option_price_integral}) is computationally difficult. In such a case, we resort to numerical techniques to approximate the option price. Algorithm~\ref{algo:solution} illustrates our Monte Carlo method. For $\widetilde{n}$ number of simulations, for each simulation, we generate a vector of multinormal noise and then calculate the CPCs at time $T$. Eq.~(\ref{eq:no_early_exercise}) shows that there is no need to generate the whole paths in each simulation as we only consider the CPCs on the expiration date in the calculation of option payoff. Hence, by having $\widetilde{n}$ payoffs at time $T$, the option price $\pi_0$ can be then approximated numerically. We refer to this as Algorithm~\ref{algo:solution}.

\begin{algorithm}[t]
\caption{Pricing a multi-keyword multi-click ad option via Monte Carlo simulations. Detailed description of notations is provided in Table~\ref{tab:notations}.}
\label{algo:solution}
\begin{algorithmic}
\Function{\texttt{OptionPricingMC}}{$\boldsymbol{K}, \boldsymbol{C}(0), \boldsymbol{\Sigma}, \boldsymbol{M}, m, r, T$}
	\For{$\; k \leftarrow 1$ to $\widetilde{n}$} \ \ \ \# $\widetilde{n}$ is the number of simulations;	
		\State $[z_{1, k}, \ldots, z_{n, k}] \leftarrow \texttt{GeneratingMultivariateNoise}(MVN[0, \boldsymbol{M \Sigma M}])$
		\For{$\; i \leftarrow 1$ to $n$}
			\State $C_{i, k} \leftarrow C_i(0) 
\exp \Big\{ (r-\frac{1}{2} \sigma_i^2) T + \sigma_i z_{i,k} \sqrt{T} \Big\}$.
		\EndFor
		\State $G_{k} \leftarrow \Phi([C_{1,k}, \ldots, C_{n, k}])$.
	\EndFor
	\State $\pi_0 \leftarrow m e^{-rT} \mathbb{E}_0 [\Phi(\boldsymbol{C}(T))] 
	\approx 
	m e^{-r T} \Big( \frac{1}{\widetilde n} \sum_{k = 1}^{\widetilde{n}} G_{k} \Big)$.	
	\State	\Return $\pi_0$
\EndFunction
\end{algorithmic}
\end{algorithm}

\subsubsection{Discussion}\label{sec:discussion}

The candidate keywords' prices may not follow the GBM assumption empirically because some time series features, such as jumps and volatility clustering, cannot be captured effectively by a GBM~\cite{Marathe_2005}. However, the GBM model is still a good choice for pricing ad options in sponsored search. First, in our data analysis (see Section~\ref{sec:gbm_test}), we find that 15.73\% keywords' CPCs satisfy the GBM assumption. Second, for the cases where the GBM assumption is not valid empirically (see Section~\ref{sec:fairness}), we find that the pricing model is reasonably robust as the identified arbitrage values in many experimental groups are small. Of course, our dataset might be biased. However, other previous research in keyword auctions support the GBM assumption: \citet{Lahaie_2007} tested the log-normality of bids on Yahoo! search advertising data and gave the estimated distribution parameters; \citet{Ostrovsky_2011} performed experiments based on the log-normal bids on Yahoo! search advertising platform; \citet{Pin_2011} observed random bids from Microsoft Bing search platform and simulated similar bids based on the log-normal distribution. Since in these research the advertisers' bids are tested across auctions, the winning payment prices (i.e., the second-highest bids from auctions) over time also satisfy the log-normal distribution. Recall that in the GBM model, the difference between two logarithms of winning payment prices follows a time dependent normal distribution. If we consider the average daily winning payment price as the underlying variable, these previous work can provide the distribution hypothesis tests to support the GBM assumption in sponsored search. However, for display advertising, the GBM assumption is usually not valid empirically, and this has been recently investigated by~\citet{Chen_2014_21} and \citet{Yuan_2014}. 

Table~\ref{tab:option_comparison} shows that if only having the exactly matched $\boldsymbol{C}(T)$, we can still construct a broad match structure for the option. Similar to Eq.~(\ref{eq:option_payoff_T}), the option payoff function on time $T$ can be defined as follows 
\begin{align}
 \Phi(\boldsymbol{C}(T))  =  & \ \max \bigg\{\sum_{i = 1}^{k_1} \omega_{1i} C_{1i}(T) - F_1, \cdots, \sum_{i = 1}^{k_n} \omega_{ni} C_{ni}(T) - F_n, 0\bigg\}. \label{eq:option_payoff_T_broad_match}
\end{align}
where $\omega_{ji}$ is the probability that the $i$th broad matched keyword (i.e., the sub-phrase occurs in search queries) for the keyword $K_j$, and $k_j$ represents the number of broad matched keywords. Eq.~(\ref{eq:sde_gbm_multi}) can be still used to model the underlying CPCs' movement and the option price $\pi_0$ can be directly calculated by Algorithm~\ref{algo:solution}.

\subsection{Revenue Analysis for Search Engine}\label{sec:effects_to_reve_1}

The proposed ad option can be considered as an \lq\lq{}insurance\rq\rq{} for an advertiser. The advertiser needs to pay the upfront option price, which contributes to the search engine's revenue. In the following discussion, we analyse the effect of an ad option on the search engine's revenue. We provide a functional analysis for the $1$-keyword $1$-click ad option in this section and leave the empirical investigation of the $n$-keyword cases to Section~\ref{sec:experiment}. 

Let $D(F)$ be the difference between the expected revenue from an ad option and the expected revenue from only keyword auctions, we then have 
\begin{align}
D(F) 
= & \ \bigg( \underbrace{C(0) \mathscr{N}[\zeta_1] - e^{-r T} F \mathscr{N}[\zeta_2] 
+ e^{-r T} F \bigg) \mathbb{P}(\mathbb{E}^{\mathbb{Q}}_0[C(T)] \geq F)}_{
= \ \textrm{Discounted value of expected revenue from option if } \mathbb{E}^{\mathbb{Q}}_0[C(T)] \geq F}  \nonumber \\[0.12in]
& \ +  \bigg( \underbrace{C(0) \mathscr{N}[\zeta_1] - e^{-r T} F \mathscr{N}[\zeta_2] 
+ e^{-r T} \mathbb{E}^{\mathbb{Q}}_0[C(T)] \bigg) \mathbb{P}(\mathbb{E}^{\mathbb{Q}}_0[C(T)] < F)}_{
= \ \textrm{Discounted value of expected revenue from option if } \mathbb{E}^{\mathbb{Q}}_0[C(T)] < F}  \nonumber \\[0.12in]
& \ - \underbrace{e^{-r T}  \mathbb{E}^{\mathbb{Q}}_0 [C(T)] }_{= \ \textrm{Discounted value of expected revenue from auction}}. \nonumber \\[0.12in]
= & \ C(0) \mathscr{N}[\zeta_1] - e^{- r T}  F \mathscr{N}[\zeta_2] 
- e^{- r T} (\mathbb{E}^{\mathbb{Q}}_0 [C(T)] - F) \times \mathbb{P}(\mathbb{E}^{\mathbb{Q}}_0 [C(T)] \geq F), \label{eq:reve_diff_2}
\end{align}
where $\mathscr{N}[\cdot]$ represents the cumulative probability of a standard normal distribution.

Let us consider the boundary values first. If $F = 0$, the option price $\pi_0$ achieves its maximum value $e^{-r T} \mathbb{E}^{\mathbb{Q}}_0[C(T)]$; therefore, $D(F) \rightarrow 0$. If $\pi_0 = 0$, the fixed CPC $F$ is as large as possible, and $\mathbb{P}(\mathbb{E}^{\mathbb{Q}}_0[C(T)] \geq F) \rightarrow 0$ and $D(F) \rightarrow 0$. Since 
\[
\ln\{C(T)/C(0)\} \sim N\big( (r-\sigma^2/2)T, \sigma^2 T \big),
\] 
we can have
\begin{align}
\mathbb{P}(\mathbb{E}^{\mathbb{Q}}_0[C(T)] \geq F) 
   = & \ \mathbb{P} \bigg( C(0) \exp \{ (r- \frac{1}{2}\sigma^2) T + \frac{1}{2} \sigma^2 T\} \geq F \bigg) \nonumber \\
   = & \ \mathbb{P} \bigg( \ln\{C(T)/C(0)\} + \ln\{F /C(T)\} - r T \leq 0 \bigg) \nonumber \\
   = & \ \mathbb{P} \bigg( \frac{1}{\sigma \sqrt{T}} \Big( \ln\{C(T)/C(0)\} - (r-\frac{1}{2}\sigma^2) T  \Big) \nonumber \\
     & \ \ \ \ \
     \leq \frac{1}{\sigma \sqrt{T}} \Big( \ln\{C(0)/F\} + r T + \sigma W(T) \Big) \bigg) = \mathscr{N}[\zeta^*], \label{eq:reve_delta_results_prob}
\end{align}
where $\zeta^* = \frac{1}{\sigma \sqrt{T}} \big( \ln\{C(0)/F\} + r T + \sigma W(T) \big) = \zeta_1 - (\frac{1}{2}\sigma \sqrt{T} - \frac{1}{\sqrt{T}} W(T)) = \zeta_2 + (\frac{1}{2}\sigma \sqrt{T} + \frac{1}{\sqrt{T}} W(T))$. Since $\mathbb{E}[W(T)] = 0$ and $\textrm{var}[W(T)] = T$, then $\zeta_2 \leq \zeta^* \leq \zeta_1$ with $67\%$ probability if $\sigma^2 T \geq 4$. Then
\begin{align}
D(F) = & \ C(0) \mathscr{N}[\zeta_1] - e^{-r T} F \mathscr{N}[\zeta_2] - e^{-r T} (\mathbb{E}^{\mathbb{Q}}_0 [C(T)] - F) \mathscr{N}[\zeta^*] \nonumber \\
=  & \ C(0) \Big( \mathscr{N}[\zeta_1] - \mathscr{N}[\zeta^*] \Big) + e^{- r T} F \Big( \mathscr{N}[\zeta^*] - \mathscr{N}[\zeta_2] \Big) \geq 0, 
\end{align}
suggesting that the search engine can have an increased expected revenue if the search engines sells the click via an option rather than through an auction. Taking the derivative of $D(F)$ with respect to $F$ and assigning its value to zero, we have
\begin{align}
\frac{\partial D(F) }{\partial F} = & \ C(0) \frac{\partial \mathscr{N}[\zeta_1]}{\partial \zeta_1} \frac{\partial \zeta_1}{\partial F} - e^{-r T} \mathscr{N}[\zeta_2] - e^{-r T} F \frac{\partial \mathscr{N}[\zeta_2]}{\partial \zeta_2} \frac{\partial \zeta_2}{\partial F} \nonumber \\
& - e^{-r T} (\mathbb{E}^{\mathbb{Q}}_0[C(T)] - F) \frac{\partial \mathbb{P}(\mathbb{E}^{\mathbb{Q}}_0[C(T)] \geq F)}{\partial F}  
+  e^{-r T} \mathbb{P}(\mathbb{E}^{\mathbb{Q}}_0[C(T)] \geq F) = 0. \label{eq:reve_diff_d}
\end{align}

Since $\partial \mathscr{N}(x)/\partial x = \frac{1}{\sqrt{2 \pi}} e^{- \frac{1}{2} x^2 }$, the following equation holds
\begin{align}
\frac{\partial \mathscr{N}[\zeta_2]}{\partial \zeta_2} 
\bigg/ 
\frac{\partial \mathscr{N}[\zeta_1]}{\partial \zeta_1} 
= & \ \exp\bigg\{\frac{1}{2}(\zeta_1^2 - \zeta_2^2)\bigg\} = \frac{C(0) e^{r T} }{F}. \label{eq:reve_cond_1}
\end{align}

Taking the derivative of $\zeta_1$ and $\zeta_2$ with respect to $F$ gives
\begin{align}
\frac{\partial \zeta_1}{\partial F} = & \ \frac{\partial \frac{1}{\sigma \sqrt{T}}
\bigg( \ln\{C(0)/F\} + (r + \frac{1}{2}\sigma^2) T \bigg) }{\partial F} = - \frac{1}{F \sigma \sqrt{T}}, \label{eq:reve_cond_2}\\
\frac{\partial \zeta_2}{\partial F} = & \ \frac{\partial \zeta_1}{\partial F} 
- \frac{\partial \sigma \sqrt{T}}{\partial F} 
= - \frac{1}{F \sigma \sqrt{T}}. \label{eq:reve_cond_3}
\end{align}
and $D(F)$ achieves its maximum or minimum value at $F = \mathbb{E}^{\mathbb{Q}}_0[C(T)]$. Further, taking the second derivative of $D(F)$ with respect to $F = \mathbb{E}^{\mathbb{Q}}_0[C(T)]$ gives
\begin{align}
\frac{\partial^2 D(F) }{\partial F^2} =  & \ \frac{\partial \mathbb{P}(\mathbb{E}^{\mathbb{Q}}_0[C(T)] \geq F)}{\partial F} 
= \frac{\partial \mathscr{N}[\zeta_2]}{\partial \zeta_2}  \frac{\partial \zeta_2}{\partial F}  =  - \frac{1}{\sqrt{2 \pi_0}} e^{- \frac{1}{2} {\zeta_2}^2 } \frac{1}{F \sigma \sqrt{T}} < 0. \nonumber
\end{align}
Hence, if the fixed CPC is set the same as the estimated spot CPC on the contract expiration date (i.e., $F = \mathbb{E}^{\mathbb{Q}}_0[C(T)]$), the search engine can increase its profit. 

\section{Experiments}\label{sec:experiment}

In this section, we describe our data and experimental settings, conduct assumption and fairness tests, and investigate the option\rq{}s effects on the search engine's revenue.

\subsection{Data and Experimental Design}\label{sec:exp_design}

\begin{table}[tp]
\centering
\tbl{Overview of experimental settings of data.\label{tab:exp_setting}}{
\begin{tabular}{|c|c|c|c|}
\hline
\hspace{-7pt} Market \hspace{-7pt} & \hspace{-7pt} Group \hspace{-7pt} & \hspace{-7pt} Training set (31 days) \hspace{-7pt} & \hspace{-3pt} Deve\&test set (31 days) \hspace{-7pt}\\
\hline
\multirow{4}{*}{US}  & 1 & 25/01/2012-24/02/2012 & 24/02/2012-25/03/2012 \\
                     & 2 & 30/03/2012-29/04/2012 & 29/04/2012-31/05/2012 \\
                     & 3 & 10/06/2012-12/07/2012 & 12/07/2012-17/08/2012 \\
                     & 4 & 10/11/2012-11/12/2012 & 11/12/2012-10/01/2013 \\                    
\hline
\multirow{4}{*}{UK}  & 1 & 25/01/2012-24/02/2012 & 24/02/2012-25/03/2012 \\
                     & 2 & 30/03/2012-29/04/2012 & 29/04/2012-31/05/2012 \\
                     & 3 & 12/06/2012-13/07/2012 & 13/07/2012-19/08/2012 \\
                     & 4 & 18/10/2012-22/11/2012 & 22/11/2012-24/12/2012 \\                    
\hline
\end{tabular}
}
\end{table}

The data used in the experiments is collected from Google AdWords by using its Traffic Estimation Service~\cite{Yuan_2012}: when an advertiser submits his targeted keywords, budget, and other settings to Google, the Traffic Estimation Service will return a list of data values, including the estimated CPCs, clicks, global impressions, local impressions and position. These values are recorded for the period from 26/11/2011 to 14/01/2013, for a total of 557 keywords in the US and UK markets. Note that in the data 21 keywords have missing values and 115 keywords's CPCs are all 0. 

For each market, as illustrated in Table~\ref{tab:exp_setting}, we split the data into 4 experimental groups and each group has one training, one development, and one test set.  The training set is used to: (i) select the keywords with non-zero CPCs; (ii) test the statistical properties of the underlying dynamic and estimate the model parameters. We then price ad options and simulate the corresponding buying and selling transactions in the development set. Finally, the test set is used as the baseline to examine the priced ad options.

\subsection{Parameter Estimation and Option Pricing}\label{sec:exp_para_estimation}

The GBM parameters are estimated by using the method suggested by~\citet{Wilmott_2006_1}. Specifically, for the keyword $K_i$, the volatility $\sigma_i$ is the sample standard deviation of change rates of log CPCs and the correlation $\rho_{i j}$ is given by
\begin{equation}\label{eq:correlation}
\rho_{i j} = \frac{\sum_{k=1}^{\widetilde{m}} \big(y_i (k) - \bar y_i 
\big)\big(y_j (k) - \bar y_j \big)}
{\sqrt{\sum_{k=1}^{\widetilde{m}} \big(y_i (k) - \bar y_i\big)^2 
\sum_{k=1}^{\widetilde{m}}  \big(y_j (k) - \bar y_j\big)^2}},
\end{equation}
where $\widetilde{m}$ is the size of training data and $y_i(t_k)$ is the $k$th 
change rate of log CPCs. 

\begin{figure}[tp]
\centering
\includegraphics[width=0.875\linewidth]{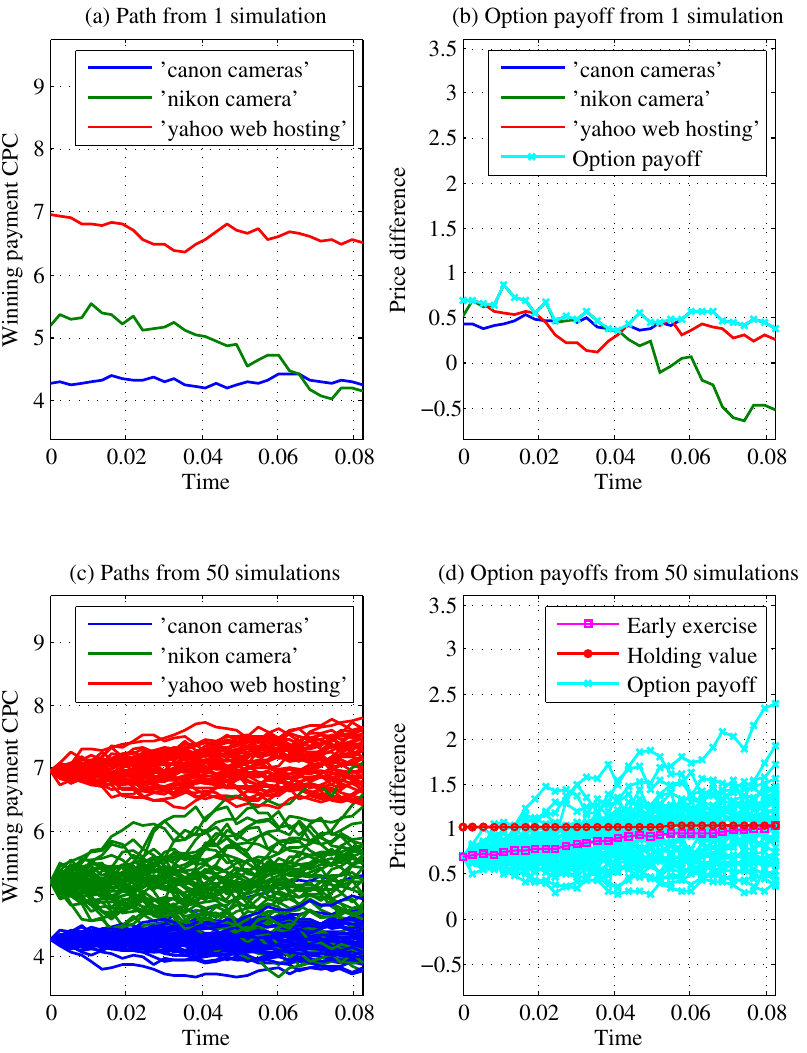}
\caption{Empirical example of pricing a 3-keyword 1-click ad option via Monte Carlo method, where $K_1 = \textrm{\lq{}canon cameras\rq{}}$, $K_2 = \textrm{\lq{}nikon camera\rq{}}$, $K_3 = \textrm{\lq{}yahoo web hosting\rq{}}$, $F_1 = 3.8505$, $F_2 = 4.6704$ and $F_3 = 6.2520$.}
\label{fig:gbm_simulation_ex}
\end{figure}

Figure~\ref{fig:gbm_simulation_ex} illustrates an 
empirical example, where the candidate keywords are
\[
\boldsymbol{K}
=
\left\{
\begin{array}{c}
	K_1 \\
    K_2 \\
    K_3 \\
\end{array}  
\right\}
= 
\left\{
\begin{array}{c}
	\textrm{\lq{}canon cameras\rq{}}\\
    \textrm{\lq{}nikon camera\rq{}}\\
    \textrm{\lq{}yahoo web hosting\rq{}}\\
\end{array}  
\right\},
\]
and the estimated model parameters are 
\[
\boldsymbol{\sigma} = 
\left(
\begin{array}{c}
	0.2263\\
    0.4521\\
    0.2136\\
\end{array}  
\right), \  
\boldsymbol{\Sigma} = 
\left( \begin{array}{ccc}
	1.0000  &  0.2341  &  0.0242\\
    0.2341  &  1.0000  & -0.0540\\
    0.0242  & -0.0540  &  1.0000\\
    \end{array} 
\right).
\]
Note that a high contextual relevance of keywords normally means that they have a high substitutional degree to each other, such as \lq{}canon cameras\rq{} and \lq{}nikon camera\rq{}, whose CPCs move in the same direction with correlation $0.2341$. The other keyword \lq{}yahoo web hosting\rq{} is contextually less relevant to the formers and also has very low price correlations to them. The example also shows that the contextual relevance of keywords has an impact on their CPCs movement. 

Based on the estimated parameters, we draw a sample of simulated paths of a 3-dimensional GBM in Figure~\ref{fig:gbm_simulation_ex}(a) for 31 days (where the x-axis is expressed in terms of year value). Recall that the option payoff at any time $t$ in the contract lifetime is $\max\{C_1(t)-F_1, \ldots, C_n(t) - F_n, 0\}$. In Figure~\ref{fig:gbm_simulation_ex}(b), we plot the price difference between the spot CPC and the fixed CPC of each candidate keyword (i.e., $C_i(t)-F_i$, $i=1,\ldots, n$) and also indicate the corresponding option daily payoffs (shown by the cyan curve). It suggests that switching among keywords would help the advertiser to maximise the benefits of the ad option. Repeating the above simulations 50 times generates 50 simulated vales of each keyword for each day, as shown in Figure~\ref{fig:gbm_simulation_ex}(c). We then calculate 50 option payoffs and their daily mean values to obtain the final option price, as shown in Figure~\ref{fig:gbm_simulation_ex}(d).

To examine the fairness (i.e., no-arbitrage) of the calculated option price, we can construct a risk-less value difference process by delta hedging $\partial V/\partial C_j$ (see Appendix~\ref{app:derivation_bsm}) and check if any arbitrage 
exists~\cite{Wilmott_2006_1}. The hedging delta of the 1-keyword 1-click ad option can be calculated as follows
\begin{equation}\label{eq:delta_hedging}
\frac{\partial V}{\partial C}=\mathscr{N} \bigg[ \frac{1}{\sigma 
\sqrt{T}} \bigg(\ln\bigg\{ \frac{C(0)}{F} \bigg\} + (r + \frac{\sigma^2}{2}) T 
\bigg)\bigg].
\end{equation}

For the $n$-keyword $1$-click option, the hedging delta of each 
keyword can be computed by the Monte Carlo method, i.e., $\partial 
V/\partial C_i = \mathbb{E}^{\mathbb Q}[\partial V(T, C(T))/\partial C_i(T)]$. According to Appendix~\ref{app:derivation_bsm}, we can define the 31-day growth rate of the value difference process as $\widetilde{\gamma} = \Big(\Pi(t_{31}) - \Pi(t_0) \Big)/\Pi(t_0)$, and compare $\widetilde{\gamma}$ to the risk-less bank interest rate $r=5\%$ (equivalent to $\widetilde{r} = 4.12\%$ per 31 days return\footnote{The relationship between the continuous compounding $r$ and the return per 31 days $\widetilde{r}$ is: $1 + \widetilde{r} = e^{r \times 30/365}$~\cite{Hull_2009}.
}). The arbitrage detection criteria is
\begin{equation}
| \widetilde{\gamma} - \widetilde{r} | \leq \varepsilon \ ? \ \textbf{arbitrage does not exist : arbitrage exists},
\end{equation}
where the notation $\varepsilon$ is the model variation threshold (and we set $\varepsilon = 5\%$ in experiments). Hence, a positive $\widetilde{\gamma} - \widetilde{r}$ means that the advertiser buys an option can obtain arbitrage while a negative $\widetilde{\gamma} - \widetilde{r}$ indicates the case of making arbitrage by selling an option. Then the identified arbitrage $\alpha$ is defined as the excess return, that is
\begin{equation}
\alpha = 
\bigg\{
\begin{array}{cc}
\widetilde{\gamma} - (\widetilde{r} - \varepsilon), & \textrm{if } \widetilde{\gamma} < \widetilde{r} - \varepsilon, \\
\widetilde{\gamma} - (\widetilde{r} + \varepsilon), & \textrm{if } \widetilde{\gamma} > \widetilde{r} + \varepsilon. 
\end{array}
\label{eq:arbitrage_exp}
\end{equation}

\begin{table}
\tbl{Test of arbitrage for ad options based on a GBM: $n$ is the number of candidate keywords, $N$ is the number of options priced in a group, $\mathbb{P}(\alpha)$ is percentage of options in a group with identified arbitrage, and the $\mathbb{E}[\alpha]$ is the average arbitrage value of the options, where the arbitrage $\alpha$ is defined by Eq.~(\ref{eq:arbitrage_exp}) and the risk-less bank interest rate $r=5\%$.
\label{tab:delta_heding_arbitrage_gbm}}{
\begin{tabular}{|c|c|c|c|c|c|c|c|}
\hline
\multirow{2}{*}{$n$} & \multirow{2}{*}{Group} & \multicolumn{3}{c|}{US market} & \multicolumn{3}{c|}{UK market}\\
\cline{3-8}
   &   &  $N$ & $\mathbb{P}(\alpha)$ & \hspace{-7pt} $\mathbb{E}[\alpha]$ \hspace{-7pt} & $N$ & $\mathbb{P}(\alpha)$ & \hspace{-7pt} $\mathbb{E}[\alpha]$ \hspace{-7pt}\\  
\hline
\multirow{4}{*}{1}   
   &   1   &  94  & 0.00\% &  0.00\%  & 76 & 0.00\% & 0.00\%\\
   &   2   &  64  & 0.00\% &  0.00\%  & 45 & 0.00\% & 0.00\%\\
   &   3   &  94  & 1.06\% &  0.75\%  & 87 & 0.00\% & 0.00\%\\
   &   4   &  112 & 0.89\% & -0.37\%  & 53 & 0.00\% & 0.00\%\\
\hline
\multirow{4}{*}{2}   
   &   1   &  47 & 4.26\% &  1.63\%  & 38 & 0.00\%  & 0.00\% \\
   &   2   &  32 & 9.38\% &  0.42\%  & 22 & 4.55\% & 13.41\% \\
   &   3   &  47 & 4.26\% &  0.84\%  & 43 & 4.65\%  & 0.82\% \\
   &   4   &  56 & 5.36\% &  3.44\%  & 26 & 23.08\% & -6.22\% \\
\hline
\multirow{4}{*}{3}   
   &   1   &  31 & 0.00\%  & 0.00\%   & 25 & 4.00\%  &  0.00\% \\
   &   2   &  21 & 4.76\%  & -1.38\%  & 15 & 0.00\%  &  0.00\% \\
   &   3   &  31 & 0.00\%  & 0.00\%   & 29 & 3.45\%  & -1.12\% \\
   &   4   &  37 & 10.81\% & 3.87\%   & 17 & 35.29\% & -2.54\% \\
\hline
\end{tabular}
}
\end{table}

\begin{figure}[tp]
\centering
\includegraphics[width=1\linewidth]{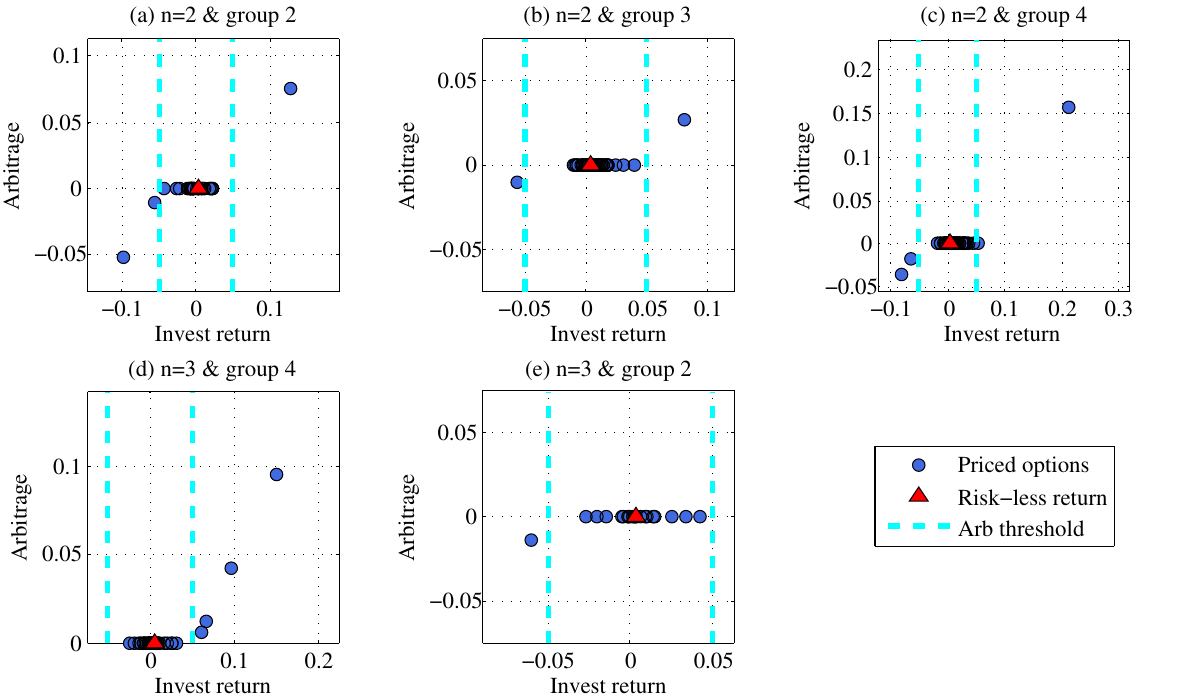}
\caption{Empirical example of arbitrage analysis based on GBM for the US market.}
\label{fig:arb_analysis_gbm_ex}
\end{figure}

Table~\ref{tab:delta_heding_arbitrage_gbm} presents the overall results of our arbitrage test based on the GBM model. We generate paths for candidate keywords with 100 simulations and examine the options price using delta hedging. There are 99.76\% (1-keyword), 93.06\% (2-keyword) and 92.71\% (3-keyword) options fairly priced. Only a small number of options exhibits arbitrage and most of the mean arbitrage values lie within 5\%, such as shown in Figure~\ref{fig:arb_analysis_gbm_ex}. The existence of small arbitrage may be due to two reasons. First, the stability of process simulations in both option pricing and arbitrage test. Second, the candidate keywords are randomly selected for the 2-keyword and 3-keyword options. The significant differences on the absolute prices of these keywords can generates a large variation of calculated option payoffs, which then trigger arbitrage. 

\subsection{Model Validation and Robustness Test}
\label{sec:model_validation}

We now examine the GBM assumption and investigate if arbitrage exists when the candidate keywords in an option do not follow a GBM.

\subsubsection{Checking the Underlying GBM Assumption}
\label{sec:gbm_test}

\begin{figure}[tp]
\centering
\includegraphics[width=1\linewidth]{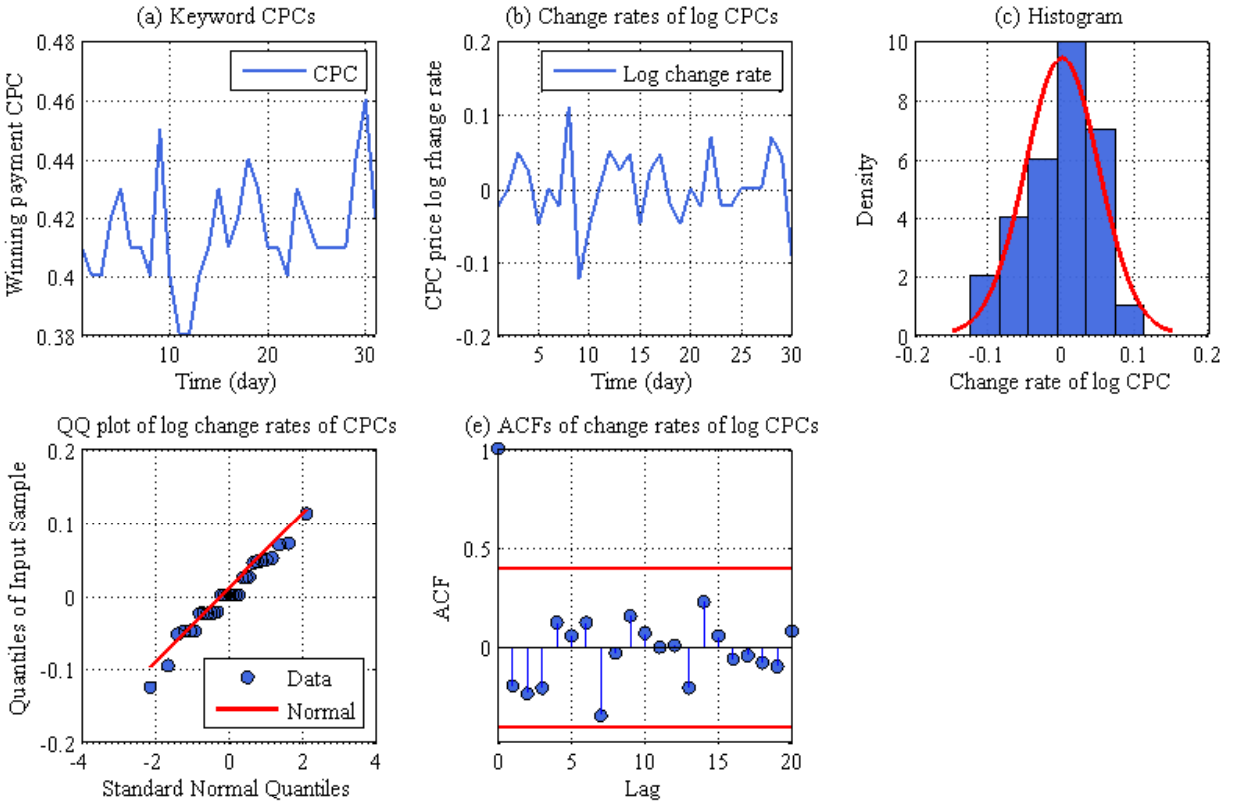}
\caption{Empirical example of checking the GBM assumption for the keyword \lq{}canon 5d\rq{}, where the Shapiro-Wilk test is with $p$-value 0.3712 and the Ljung-Box test is with $p$-value 0.4555.}
\label{fig:exp_gbm_assumption_check_ex}
\vspace{10pt}
\includegraphics[width=0.75\linewidth]{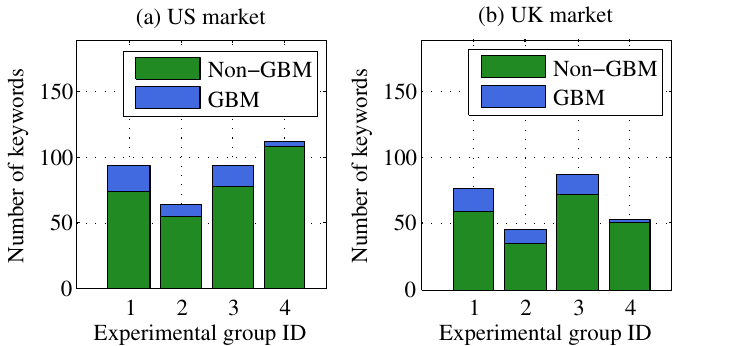}
\caption{Overview of checking the GBM assumption for all keywords of experimental groups.}
\label{fig:exp_gbm_assumption_check_overview_ex}
\end{figure}

Two validation conditions of the GBM model are tested~\cite{Marathe_2005}: (i) the normality of change rates of log CPCs; and (ii) the independence from previous data. Normality can be either checked graphically by histogram/Q-Q plot or verified statistically by the Shapiro-Wilk test~\cite{Shapiro_1965}. To examine independence, we employ the autocorrelation function (ACF)~\cite{Tsay_2005} and the Ljung-Box statistic~\cite{Ljung_1978}. Figure~\ref{fig:exp_gbm_assumption_check_ex} provides an empirical example of the keyword \lq{}canon 5d\rq{}. Figure~\ref{fig:exp_gbm_assumption_check_ex} (a)-(b) exhibit the movement of CPCs and log change rates while Figure~\ref{fig:exp_gbm_assumption_check_ex} (c)-(d) show that the stated two conditions are satisfied in this case. 

We check the discussed two conditions with the training data. As shown in Figure~\ref{fig:exp_gbm_assumption_check_overview_ex}, there are 14.25\% and 17.20\% of keywords in US and UK markets that satisfy the GBM assumption, respectively. Thus 15.73\% of keywords can be effectively priced into an option based on a GBM. It is worth mentioning that not all keywords follow a GBM. Next, we examine the robustness of pricing model and investigate the arbitrage based on non-GMB models.

\subsubsection{Examining Arbitrage for Non-GBM Dynamics}
\label{sec:fairness}

\begin{figure}[tp]
\centering
\includegraphics[width=0.875\linewidth]{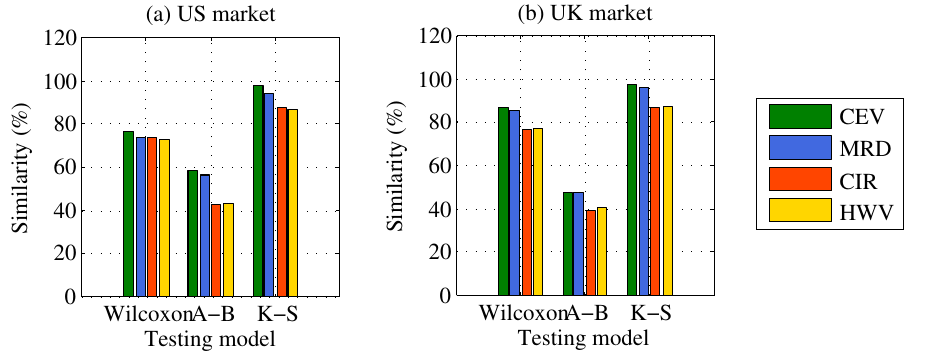}\vspace{5pt}
\caption{Overview of model similarity tests: Wilcoxon test, Ansari-Bradley (A-B) test and Two-sample Kolmogorov-Smirnov (K-S) test.}
\label{fig:model_fitness}
\vspace{10pt}
\includegraphics[width=0.75\linewidth]{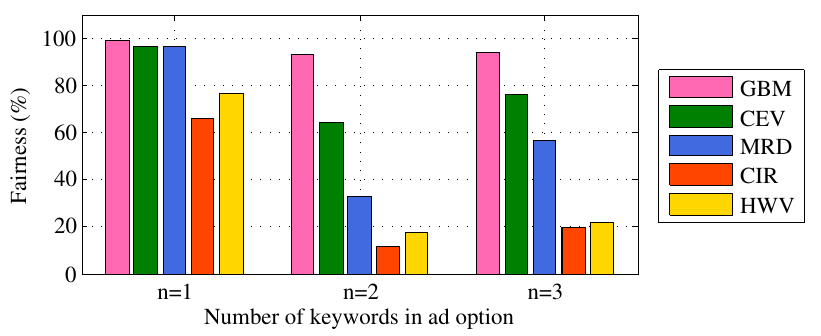}
\caption{Overview of pricing model robust tests.}
\label{fig:option_pricing_overview}
\end{figure}

Several popular stochastic processes (together with the real data) are tested to check the arbitrage in option pricing. Table~\ref{tab:non_gbm_dynamics} shows the candidate models and each model can capture certain features of time series data, such as mean-reversion, constant volatility and square root volatility~\cite{Hull_2009}. The arbitrage tests here are slightly different from that of the GBM model. We estimate the model parameters from the actual data in the test sets instead of the learning sets and treat the actual data as one single path of each model. Hence, the simulated data has the same drift, volatility and correlations as the test data. We are now able to examine the arbitrage multiple times when the real-world environment does not follow a GBM. Also, for the candidate models, hypothesis tests are used to check if the simulated path and actual data come from a same distribution. These tests include the Wilcoxon test~\cite{Frank_1945}, Ansari-Bradley test~\cite{Mood_1974} and Two-sample Kolmogorov-Smirnov test~\cite{Justel_1997}. Figure~\ref{fig:model_fitness} summarises the results of models' goodness-of-fit tests, where the y-axis represents the mean percentage of simulated paths not rejected by the hypothesis tests. Even though the three tests give different absolute percentages, the dynamics' performance is similar and consistent: the CEV model has the best simulations for the actual data, followed by the MRD; the CIR and HWV models are very close.

Table~\ref{tab:delta_heding_arbitrage_nongbm} presents the results of arbitrage tests for the non-GBM dynamics, where most of experimental groups exhibit arbitrage. The CEV model gives the best no-arbitrage performance, showing that 78.65\% of CEV-based keywords can be properly priced by using the GBM-based option pricing model. About 53.05\% of the CIR model and about 43\% of the MRD or HWV models based options have no arbitrage. For 1-keyword options, the fairness percentage is more than 85\% across all experimental groups. However, this number drops to around 38\% for multi-keyword options (36.27\% for 2-keyword options and 42\% for 3-keyword options). For the identified arbitrage, many groups (especially 1-keyword options) show small arbitrage values (around 10\%) while arbitrage 
explodes in some groups.

\begin{sidewaystable}
\tbl{Tested non-GBM dynamics: $k_i = 0.5$ while other parameters are learned 
from the training data.\label{tab:non_gbm_dynamics}}{
\begin{tabular}{|c|c|}
\hline
Dynamic & Stochastic differential equation (SDE)\\
\hline
Constant elasticity of variance (CEV) model~\cite{Cox_1976} & $ d C_i (t) = 
\mu_i C_i (t) dt + \sigma_i (C_i (t))^{1/2} d W_i (t)$ \\
Mean-reverting drift (MRD) model~\cite{Wilmott_2006_1} & $ d C_i (t) 
= k_i (\mu_i - C_i (t) )dt + \sigma_i (C_i (t))^{1/2} d W_i (t)$ \\
Cox-Ingersoll-Ross (CIR) model~\cite{Cox_1985} & $ d 
C_i (t) = k_i (\mu_i - C_i (t) )dt + (\sigma_i)^{1/2} C_i (t) d W_i (t)$ \\
Hull-White/Vasicek (HWV) model~\cite{Hull_1990} & $ d C_i (t) = k_i (\mu_i - 
C_i (t) )dt + \sigma_i d W_i (t)$ \\
\hline
\end{tabular}
}
\vspace{25pt}
\tbl{Overview of delta hedging arbitrage testing for non-GBM dynamics: same notations as in Table~\ref{tab:delta_heding_arbitrage_gbm}.
\label{tab:delta_heding_arbitrage_nongbm}}{
\begin{tabular}{|c|c|c|c|c|c|c|c|c|c|c|c|}
\hline
\multirow{2}{*}{Market} & \multirow{2}{*}{$n$} & \multirow{2}{*}{Group} & \multirow{2}{*}{$N$} & \multicolumn{2}{c|}{real data + CEV simu} & \multicolumn{2}{c|}{real data 
+ MRD simu} & \multicolumn{2}{c|}{real data + CIR simu} & 
\multicolumn{2}{c|}{real data + HWV simu}\\
\cline{5-12}
   &   &  & 
   & $\mathbb{P}(\textrm{arb})$ & \hspace{-7pt} Mean arb \hspace{-7pt}
   & $\mathbb{P}(\textrm{arb})$ & \hspace{-7pt} Mean arb \hspace{-7pt}
   & $\mathbb{P}(\textrm{arb})$ & \hspace{-7pt} Mean arb \hspace{-7pt}
   & $\mathbb{P}(\textrm{arb})$ & \hspace{-7pt} Mean arb \hspace{-7pt} \\
\hline
\multirow{12}{*}{US}
& 
\multirow{4}{*}{1} 
   &   1   & 74  & 2.70\% & -1.97\%    &  8.11\% & -0.38\%  & 75.68\% &  1.94\% 
 & 56.76\% & -0.01\% \\
&   &   2   & 55  & 0.00\% &  0.00\%    &  1.82\% & 1.80\%   & 16.36\% &  1.86\% 
 &  9.09\% & -0.93\% \\          
&   &   3   & 77  & 9.09\% &  -10.26\% &  5.19\% & -7.93\%  & 42.86\% &  0.85\%  
& 23.38\% & -1.51\% \\          
&   &   4   & 108 & 3.70\% &  1.38\%   &  2.78\% &  4.27\%  &  7.41\% &   3.99\% 
 &   8.33\% &  2.84\% \\           
\cline{2-12}
&
\multirow{4}{*}{2} 
   &   1   & 37  & 24.32\% &  3.63\%    &  81.08\% & -4.09\%  & 97.30\% &  
-16.24\%  & 97.30\% & -13.24\% \\
&   &   2   & 27  & 37.04\% &  6.01\%    &  70.37\% &  5.36\%   & 85.19\% &  
11.01\%  &  85.19\% &  10.51\% \\          
&   &   3   & 38  & 31.58\% &  5.97\%    &  31.58\% & -0.41\%  & 73.68\% &  
-6.91\%   &  57.89\% & -5.96\% \\          
&   &   4   & 54  & 29.63\% &  5.95\%    &  81.48\% &  6.61\%  &  94.44\% &  
16.78\%  &  94.44\% &  16.25\% \\          
\cline{2-12}
&
\multirow{4}{*}{3} 
   &   1   & 24 & 45.83\% & -1.36\%  &  79.17\% & -6.04\% & 100.00\% &  
-19.99\%  & 100.00\% & -17.44\% \\
&   &   2   & 18 & 11.11\% & -2.00\%  &  22.22\% & -5.11\% & 55.56\% & -4.39\% & 
 72.22\% & -2.22\% \\          
&   &   3   & 25 & 24.00\% &  7.01\%  &  32.00\% & -3.71\% & 84.00\% & -11.14\% 
& 76.00\% & -10.26\% \\          
&   &   4   & 36 & 16.67\% &  3.91\%  &  30.56\% &  2.66\% & 83.33\% &  2.73\% & 
 88.89\% &  3.38\% \\          
\hline
\multirow{12}{*}{UK}
& 
\multirow{4}{*}{1} 
   &   1   & 58  & 0.00\% &  0.00\%    &  0.00\% &  0.00\%   & 74.14\% &  
1.95\%  & 55.17\% & -1.60\% \\
&   &   2   & 35  & 0.00\% &  0.00\%    &  2.86\% & 1.51\%    & 22.86\% &  
1.65\%  & 14.29\% &  2.29\% \\          
&   &   3   & 72  & 5.56\% &  -5.78\%   &  1.39\% & -10.38\% & 29.17\% &  1.03\% 
 & 18.06\% & -0.32\% \\          
&   &   4   & 50 &  4.00\% &  5.55\%    &  6.00\% &  4.47\%   & 10.00\% &  
4.56\%  &   8.00\% &  3.45\% \\   
\cline{2-12}
&
\multirow{4}{*}{2} 
   &   1   & 29  & 37.93\% &  -0.76\%   &  62.07\% & -5.52\%  & 89.66\% &  
-14.55\%  & 72.41\%  & -11.39\% \\
&   &   2   & 17  & 47.06\% &  2.18\%    &  82.35\% &  5.00\%   & 100.00\% &  
9.87\%  &  100.00\% &  8.62\% \\          
&   &   3   & 36  & 19.44\% &  2.71\%    &  33.33\% & -1.78\%  & 75.00\% &  
-5.24\%   &  61.11\%  & -3.58\% \\          
&   &   4   & 25  & 64.00\% &  6.71\%    &  96.00\% &  9.01\%  &  92.00\% &  
21.17\%  &  92.00\%  &  20.04\% \\        
\cline{2-12}
&
\multirow{4}{*}{3} 
   &   1   & 19 & 26.32\% & -1.56\%  &  84.21\% & -5.21\% & 100.00\% &  
-16.33\%  & 78.95\% & -16.34\% \\
&   &   2   & 11 & 18.18\% &  0.40\%  &  18.18\% & -1.09\% & 63.64\%   & -1.28\% 
&  63.64\% & -1.05\% \\          
&   &   3   & 24 & 16.67\% &  3.45\%  &  25.00\% & -1.61\% & 79.17\%   & -9.14\% 
& 66.67\% & -9.23\% \\          
&   &   4   & 16 & 37.50\% &  7.83\%  &  43.75\% &  7.37\%  & 81.25\%   &  
0.68\% &  81.25\% &  8.64\% \\          
\hline
\end{tabular}
}
\end{sidewaystable}

In summary, Tables~\ref{tab:delta_heding_arbitrage_gbm} 
and~\ref{tab:delta_heding_arbitrage_nongbm} illustrate that our option pricing methods are effective and reasonably robust for the real sponsored search data. As shown in Figure~\ref{fig:option_pricing_overview}, when the keywords's price follow a GBM (15.73\%), the pricing model ensures that 95.17\% of ad options are fairly priced under the 5\% arbitrage precision. For the non-GBM keywords, the CEV model is the best performance model, giving 78.65\% of fairness; the CIR model is worst performance model and is with only 31.97\% of fairness. Overall, the best expected fairness for all keywords 
is 81.25\% while the worst is 41.91\%. We find that the increase of the number of candidate keywords in an ad option increases the likelihood of arbitrage. This is confirmed by the fact that expected fairness drops from 86.83\% (99.76\% GBM and 83.60\% non-GBM for 1-keyword options) to 43.69\% (2-keyword options) and 53.39\% (3-keyword options), respectively.

\subsection{Effects on Search Engine's Revenue}\label{sec:effects_to_reve_2}

\begin{figure}[tp]
\centering
\includegraphics[width=1\linewidth]{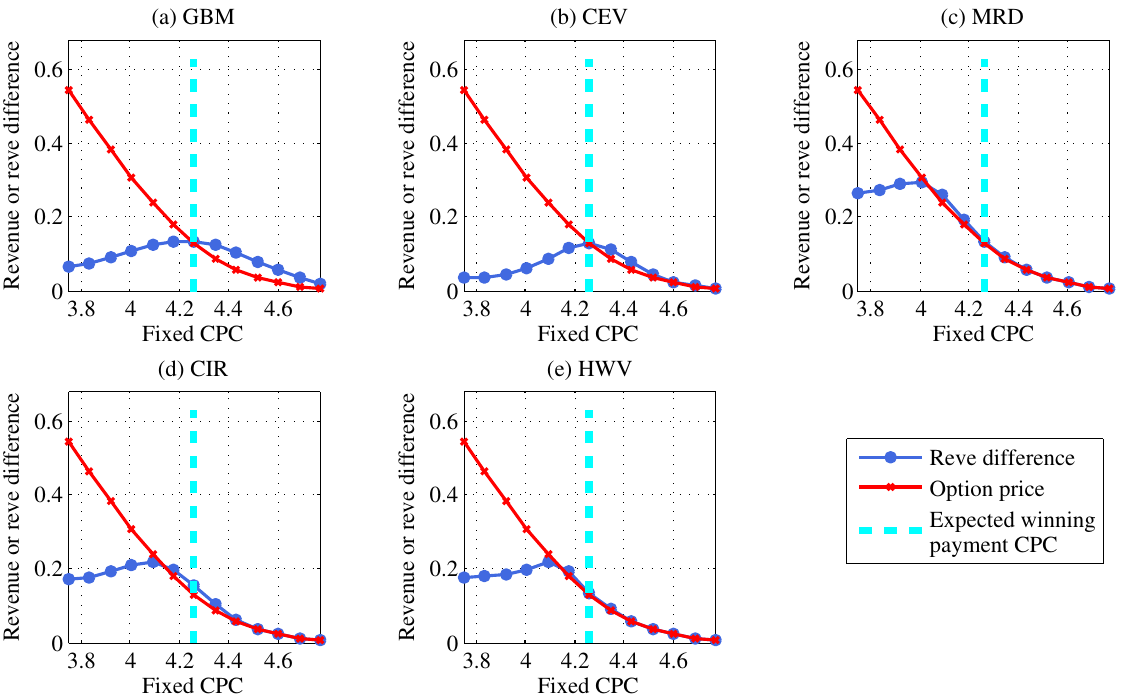}
\vspace{-5pt}
\caption{Empirical example of analysing the search engine's revenue for the keyword \lq{}canon cameras\rq{}.}
\label{fig:reve_diff_1}
\end{figure}

Let us start with the case of 1-keyword options. The example of keyword \lq{}canon cameras\rq{} in Figure~\ref{fig:reve_diff_1}(a) illustrates (other keywords exhibit the similar pattern) the conclusions from our theoretical analysis in Section~\ref{sec:effects_to_reve_1} that (i) the revenue difference between option and auction is always positive and (ii) that when the fixed CPC $F = \mathbb{E}^{\mathbb{Q}}_t [C(T)]$, the revenue difference $D(F)$ achieves its maximum and the two boundary values are approximately zero.

The non-GBM cases are further examined in Figure~\ref{fig:reve_diff_1}(b)-(e), which show that when the fixed CPC is close to zero, the revenue difference $D(F) \rightarrow 0$. This is because when the fixed CPC approximates zero, it is almost certain that the option will be used in the contract period. As such, the only income for the keyword is from the option price, which in this case is close to the CPC in the auction market (discounted back to $t$=0). On the other hand, if the fixed CPC is very high, it is almost certain that the option won't be used. In this case, the option price $\pi_0 \rightarrow 0$ and the probability of exercising the option $\mathbb{P}(\mathbb{E}^{\mathbb{Q}}_t[C(T)] \geq F) \rightarrow 0$. Hence, $D(F)$ is zero. However, under the non-GBM dynamics, the point $F = \mathbb{E}^{\mathbb{Q}}_t [C(T)]$ is not the optimal value that gives the maximum $D(F)$, which indicates that arbitrage may occur.

\begin{figure}[tp]
\centering
\includegraphics[width=1\linewidth]{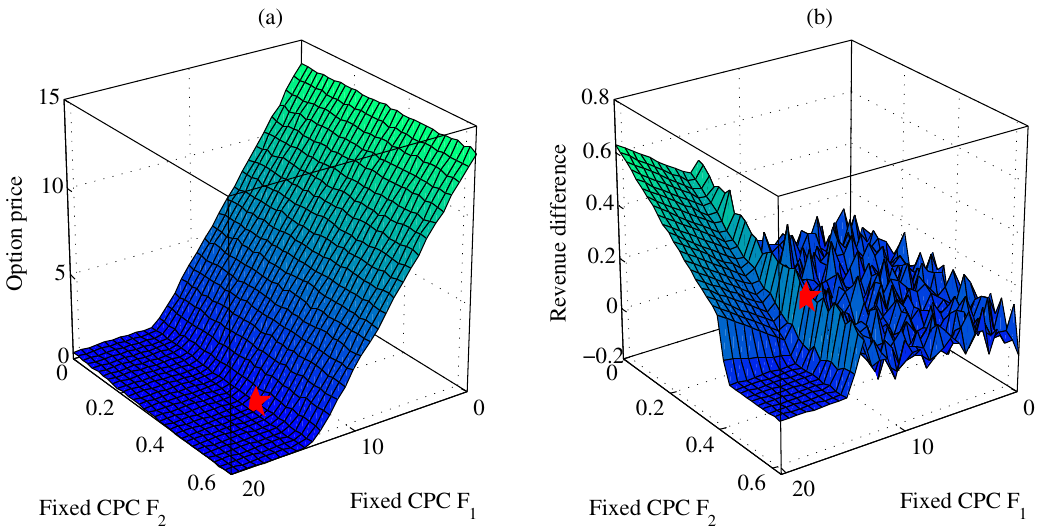}
\caption{Empirical example of analysing the search engine's revenue for the keywords \lq{}non profit debt consolidation\rq{} and \lq{}canon 5d\rq{}, where $\rho = 0.2247$.}
\label{fig:reve_diff_2}
\end{figure}

Next, Figure~\ref{fig:reve_diff_2} illustrates an empirical example a 2-keyword ad option. The candidate keywords are \lq{}non profit debt consolidation\rq{} and \lq{}canon 5d\rq{}. Figure~\ref{fig:reve_diff_2}(a) tells that the higher the fixed CPCs the lower is the option price (even though the option price is less sensitive to the keyword \lq{}canon 5d\rq{}) and it achieves the maximum when all the fixed CPCs are zeros. These monotone results are as same as the 1-keyword options. Figure~\ref{fig:reve_diff_2}(b) then shows the revenue difference curve of the search engine, where the red star represents the value where $F_1 = \mathbb{E}^{\mathbb{Q}}_t [C_1(T)]$ and $F_2 = \mathbb{E}^{\mathbb{Q}}_t [C_2(T)]$. The expected revenue differences are all non-negative, showing that this 2-keyword ad option is beneficial to the search engine's revenue. However, the red star point is not the maximum difference revenue. This is different to the 1-keyword ad options.

For higher dimensional ad options (i.e., $n \geq 3$), it is not possible to graphically examine the revenue difference. However, based on the earlier discussions, two findings can be summarised. First, there are boundary values of the revenue differences. If every $F_i \rightarrow 0$, $D(\mathbf{F}) \rightarrow 0$; and if every $F_i \rightarrow \infty$, $D(\mathbf{F}) \rightarrow 0$. Second, there exists a maximum revenue difference value even though this may not at the point where $F_i = \mathbb{E}^{\mathbb{Q}}_t [C_i(T)]$. Hence, compared to only keyword auctions, proper setting the fixed CPCs can increase the search engine's expected revenue.

\section{Concluding Remarks}\label{sec:conclusion}

In this paper, we proposed a novel framework to provide flexible guaranteed deliveries for sponsored search, from which both buy and sell sides can benefit. On the buy side, advertisers are able to secure a certain number of clicks from their targeted keywords in the future and can decide how to advertise later. They can be released from auction campaigns and can manage price risk under the given budgets. On the sell side, the search engine can sell the future clicks in advance and can receive a more stable and increased expected revenue over time. In addition, advertisers would be more loyal to a search engine due to the contractual relationships, which has the potential to boost the search engine's revenue on the long run. 

We also believe that the proposed ad options will soon be welcomed by the sponsored search market. Several similar but different developments appeared in the display digital markets are able to support our point of view. They are:
\begin{description}
\item[09/2013:] AOL's Programmatic Upfront\footnote{\url{http://www.aolplatforms.com/aolupfront2014}.}.
\item[03/2013:] OpenX Programmatic Guarantee~\cite{OpenX_2013}.
\item[10/2012:] Adslot Media's Programmatic Direct Media Buying\footnote{\url{http://www.automatedguaranteed.com}.}.
\item[10/2012:] Shiny Ads Programmatic Direct Advertising Platform\footnote{\url{https://shinyads.com/solutions/direct-programmatic-guaranteed}.}.
\item[10/2012:] iSOCKET's Programmatic Direct\footnote{\url{https://www.isocket.com/automated-guaranteed}.}.
\end{description}
Our work differs to the above developments in many aspects. First, we focus on sponsored search while they are for display advertising. Second, the proposed ad options provide flexible guaranteed deliveries (e.g., multi-keyword targeting, multi-click exercise, early exercise, no obligation of exercise) while other recent developments do not provide such features.

Our work leaves several directions for future research. First, to address the limitations of GBM, other stochastic processes tailored to some specific keywords are worth studying, such as the jump-diffusion model~\cite{Kou_2002} and the stochastic volatility model~\cite{Chen_2014_3}. The most challenging part of this future research is that the underlying model is multi-dimensional and needs to be computational fast. Second, it would be interesting to discuss an optimal pricing and allocation model of ad options so that a search engine can algorithmically manipulate the limited future clicks in front of uncertain demand. Third, the game-theoretical pricing of ad options can be another direction.

\bibliographystyle{ACM-Reference-Format-Journals}
\bibliography{mybib}

\appendix
\section*{APPENDICES}

\section{Proof of the No-Early Exercise Property for the Proposed Ad Option}
\label{app:proof_no_early_exercise}

Eq.~(\ref{eq:option_payoff_T}) can be rewritten as $\Phi(\boldsymbol{x}) = \max\{ \boldsymbol{x} - \boldsymbol{f}, 0\}$, where $\boldsymbol{x}^\prime = [x_1, \ldots, x_n]$ and $\boldsymbol{f}^\prime = [f_1, \ldots, f_n]$. It is not difficult to find that $\Phi(\boldsymbol{x})$ is multivariate convex. Let $0 \leq \lambda \leq 1$ and let $\boldsymbol{y}^\prime = [y_1, \ldots, y_n]$, if the elements of vector $\boldsymbol{a} = \boldsymbol{y} - \boldsymbol{x}$ are all non-negative, then
\begin{align}
\Phi\big(\lambda \boldsymbol{x} + (1 - \lambda) \boldsymbol{y} \big) \leq & \ \lambda \Phi(\boldsymbol{x}) + (1 - \lambda) \Phi(\boldsymbol{y}). \nonumber
\end{align}
If taking $\boldsymbol{y}^\prime = (0, \ldots, 0)$, and using the fact that $\Phi(\boldsymbol{0}) = 0$, we obtain 
\begin{equation}
\Phi(\lambda \boldsymbol{x}) \leq \lambda \Phi\big( \boldsymbol{x} \big), \ \ \ \textrm{for all } x_i \geq 0, \ 0 \leq \lambda \leq 1.
\nonumber
\end{equation}
For $0 \leq s \leq t \leq T$, since $0 \leq e^{-r (t - s)} \leq 1$, we then have
\begin{align}
\mathbb{E}^{\mathbb{Q}}_s \big[ e^{-r(t- s)} \Phi\big(\boldsymbol{X}(t)\big) \big] 
\geq  & \ \mathbb{E}^{\mathbb{Q}}_s \big[ \Phi \big(e^{-r(t- s)} \boldsymbol{X}(t) \big) \big]  \nonumber \\
\geq & \ \Phi \big( \mathbb{E}^{\mathbb{Q}}_s \big[ e^{-r(t- s)} \boldsymbol{X}(t) \big] \big) 
\ \ (\textrm{by the Jenen's Inequality}) \nonumber \\
= & \ \Phi \big(e^{rs} \mathbb{E}^{\mathbb{Q}}_s \big[ e^{-rt} \boldsymbol{X}(t) \big] \big),  \nonumber
\end{align}
where $\mathbb{E}^{\mathbb{Q}}_s[\cdot]$ is the conditional expectation with respect to time $s$ under the risk-neutral probability measure $\mathbb{Q}$. Since $e^{-rt} \boldsymbol{X}(t)$ is a martingale under $\mathbb{Q}$~\cite{Bjork_2009}, then
\begin{equation}
\Phi \big( e^{rs} \mathbb{E}^{\mathbb{Q}}_s \big[ e^{-rt} \boldsymbol{X}(t) \big] \big)  
= \Phi \big(e^{rs}  e^{-rs} \boldsymbol{X}(s) \big) = \Phi \big( \boldsymbol{X}(s) \big). \nonumber
\end{equation}
Hence, $\mathbb{E}^{\mathbb{Q}}_s \big[ e^{-r(t- s)} \Phi\big( \boldsymbol{X}(t) \big) \big] \geq \Phi \big( \boldsymbol{X}(s) \big)$, 
showing that $e^{-rt} \Phi\big(\boldsymbol{X}(t)\big)$ is a sub-martingale under $\mathbb{Q}$. This tells that the proposed ad option can be priced as same as its European structure, focusing on the payoff on the contract expiration date. For further detailed discussions about martingale and sub-martingale, please see~\citep{Bjork_2009}.

\section{Derivation of the Ad Option Pricing Formula}
\label{app:derivation_bsm}

Since the proposed ad option complements the existing keyword auctions, there may exist a situation that some advertisers only want to make guaranteed profits from the difference of costs between option and auction markets without taking any risk. This situation is called \emph{arbitrage}~\cite{Varian_1987,Bjork_2009}. Hence, we must fairly evaluate the option so that arbitrage is eliminated.

In the context of sponsored search, we consider that an advertiser buys a $n$-keyword $m$-click ad option at time $0$. Then at time $t$, $t \in [0, T]$, the difference between the option value and the market value of candidate keywords can be expressed as
\begin{equation} 
\Pi(t) = V(t, \boldsymbol{C}(t); \boldsymbol{F}, T, m) - \sum_{i = 1}^{n} \psi_i(t) C_i(t), \label{eq:cost_portfolio}
\end{equation}
where $\psi_i(t)$ represents the number of clicks needed for the keyword $K_i$ such that $\sum_i \psi_i(t)= m$. Here we call $\Pi(t)$ as the \emph{value difference process}. Recall that in Eq.~(\ref{eq:option_payoff_T}), we consider the value of an $n$-keyword $m$-click option as the sum of $m$ independent $n$-keyword $1$-click options, for the mathematical convenience, Eq.~(\ref{eq:cost_portfolio}) can be rewritten as follows
\begin{align}\label{eq:pi}
\Pi(t) =  & m \Bigg( V(t, \boldsymbol{C}(t); \boldsymbol{F}, T, 1) - \sum_{i = 1}^n \Delta_i C_i (t) \Bigg),
\end{align}
where $\Delta_i$ represents the probability that a single click goes for the keyword $K_i$ and $\sum_{i=1}^n \Delta_i = 1$. The changes of $\Pi$ over a sufficient small period of time $d t$ is then 
\begin{align}
d \Pi(t)
= & \ \ m \Bigg( 
\frac{\partial V}{\partial t} d t 
+ \frac{1}{2} \sum_{i = 1}^n \sum_{j = 1}^n \sigma_i \sigma_j \rho_{i j} C_i C_j \frac{\partial^2 V}{\partial C_i \partial C_j} d t  
+ \sum_{i = 1}^n \frac{\partial V}{\partial C_i} d C_i 
- \sum_{i = 1}^n \Delta_i d C_i 
\Bigg) . 
\end{align}
The uncertain components in $d \Pi(t)$ can be removed if $\Delta_i = \partial V/ \partial C_i$. This is called the \emph{delta hedging} in option pricing theory~\cite{Wilmott_2006_1}. Hence, $\Pi(t)$ now becomes a risk-less process over time  
\begin{align}
d \Pi(t)
= & \ m \Bigg(\frac{\partial V}{\partial t} 
+ \frac{1}{2} \sum_{i = 1}^n \sum_{j = 1}^n \sigma_i \sigma_j \rho_{i j} C_i C_j \frac{\partial^2 V}{\partial C_i \partial C_j} \Bigg) d t. \label{eq:d_pi_1}
\end{align}

We assume that the advertiser has no initial fund and he borrows the money from others at the risk-less bank interest rate $r$, so the interest of this borrowing is 
\begin{equation} 
d \Pi (t) = r \Pi (t) dt = r m \Bigg(V - \sum_{i = 1}^n \frac{\partial V}{\partial C_i} C_i\Bigg) dt. \label{eq:d_pi_2}
\end{equation} 

Eqs.~(\ref{eq:d_pi_1})-(\ref{eq:d_pi_2}) need to be equal otherwise arbitrage exists. If the risk-less growth rate of the value difference process is larger than the risk-less bank interest rate, the advertiser can obtain arbitrage by: (i) borrowing the money from bank at interest rate $r$ to buy an ad option first; (ii) selling the ad option later to repay the bank interest. In the case where the risk-less growth rate of the value difference process is smaller than the risk-less bank interest rate, the advertiser can obtain the risk-less surplus by: (i) selling short an ad option first and saving the revenue in a bank account; (ii) using the deposit money to buy the clicks of underlying keywords later. In either case, the advertiser can finally receive a risk-less surplus; therefore, arbitrage exists. 

Solving Eqs.~(\ref{eq:d_pi_1})-(\ref{eq:d_pi_2}) can give a parabolic partial differential equation (PDE) for the no-arbitrage equilibrium:
\begin{equation}\label{eq:pde_bsm}
\frac{\partial V}{\partial t}
+ r \sum_{i = 1}^n \frac{\partial V}{\partial C_i} C_i 
+ \frac{1}{2} \sum_{i = 1}^n \sum_{j = 1}^n \frac{\partial^2 V}{\partial C_i \partial C_j} \sigma_i \sigma_j \rho_{i j} C_i C_j  - r V = 0. \nonumber
\end{equation}
The above PDE satisfies the boundary condition in Eq.~(\ref{eq:option_payoff_T}). By employing the multidimensional Feynman-Ka\u{c} stochastic representation~\cite{Bjork_2009}, we obtain the solution
\begin{equation}
V(t, \boldsymbol{C}(t); \boldsymbol{F}, T, 1) 
= e^{-r(T-t)} \mathbb E^{\mathbb Q}_t [ \Phi(\boldsymbol C(T)) ], \nonumber \label{eq:pde_solution}
\end{equation}
where $\mathbb{E}^{\mathbb{Q}}_t[\cdot]$ is the conditional expectation with respect to time $t$ under the risk-neutral probability $\mathbb{Q}$. The process $C_i(t)$ can be rewritten as
\begin{align}
d C_i (t) = & \ r C_i(t) dt + \sigma_{i} C_i(t) d W_i^{\mathbb{Q}}(t), \nonumber \label{eq:sde_gbm_multi_risk_neutral}
\end{align}
where $W_i^{\mathbb{Q}}(t)$ is the standard Brownian motion under $\mathbb{Q}$. Therefore, the option price $\pi_0$ can be calculated by the following formula:
\[
\pi_0 = V(0,\boldsymbol{C}(0); \boldsymbol{F}, T, m) 
= m V(0,\boldsymbol{C}(0); \boldsymbol{F}, T, 1)
= m e^{-r T} \mathbb E^{\mathbb Q}_0 [ \Phi(\boldsymbol C(T)) ].
\]

\section{Option Pricing Formulas for Special Cases}
\label{app:special_cases}

If $n=1$, Eq.~(\ref{eq:option_price_integral}) is equivalent to the Black-Scholes-Merton (BSM) pricing formula for an European call option~\cite{Black_1973,Merton_1973}. Then we have
\begin{equation}
\pi_0 = m C(0) \mathscr{N}[\zeta_1] - m F e^{-r T} \mathscr{N}[\zeta_2], \label{eq:option_price_bsm}
\end{equation}
where $\zeta_1 = \frac{1}{\sigma \sqrt{T}} \big( \ln \{C(0)/F\} + (r+ \frac{\sigma^2}{2}) T \big)$ and $\zeta_2 = \zeta_1 - \sigma \sqrt{T}$.

If $n=2$, Eq.~(\ref{eq:option_price_integral}) contains a bivariate normal distribution. Hence, we can calculate the option price as same as the dual-strike European call option~\cite{Zhang_1998}:
\begin{align}
\pi_0 = & \ m C_1(0) \bigintssss_{-\infty}^{\zeta_1 + \sigma_1 \sqrt{T}} 
\hspace{-22pt} f(u) 
\mathscr{N}\bigg[\frac{q_1(u+\sigma_1\sqrt{T})-\rho \sigma_1 \sqrt{T}+\rho u}{\sqrt{1-\rho^2}} \bigg] du \nonumber \\
& \ + m C_2(0) \bigintssss_{-\infty}^{\zeta_2 + \sigma_2 \sqrt{T}} 
\hspace{-22pt} f(v) 
\mathscr{N}\bigg[\frac{q_2(u+\sigma_2\sqrt{T})-\rho \sigma_1 \sqrt{T}+\rho v}{\sqrt{1-\rho^2}} \bigg] dv \nonumber \\
& \ - m e^{-rT} \Bigg( F_1 \bigintssss_{-\infty}^{\zeta_1} f(u) 
\mathscr{N}\bigg[\frac{q_1(u) +\rho u}{\sqrt{1-\rho^2}} \bigg] du 
+ F_2 \bigintssss_{-\infty}^{\zeta_2} f(v) 
\mathscr{N}\bigg[\frac{q_2(v) +\rho v}{\sqrt{1-\rho^2}} \bigg] dv
\Bigg), \label{eq:option_price_dual_strike}
\end{align}
where 
\begin{align}
q_1(u) = & \ \frac{1}{\sigma_2 \sqrt{T}}
\Bigg( \ln\bigg\{\frac{F_2 - F_1 + C_1(0)e^{(r-\frac{1}{2}\sigma_1^2)T-u \sigma_1 \sqrt{T}}}{C_2(0)} \bigg\} 
- (r- \frac{1}{2}\sigma_2^2) T \Bigg),   \nonumber \\
q_2(u) = & \ \frac{1}{\sigma_1 \sqrt{T}}
\Bigg( \ln\bigg\{\frac{F_1 - F_2 + C_2(0)e^{(r-\frac{1}{2}\sigma_2^2)T-v \sigma_2 \sqrt{T}}}{C_1(0)} \bigg\} - (r- \frac{1}{2}\sigma_1^2) T \Bigg),\nonumber \\
\zeta_1 = & \ \frac{1}{\sigma_1 \sqrt{T}} \Bigg( \ln \{C_1(0)/F_1\} + (r - \frac{1}{2} \sigma_1^2) T \Bigg),  \nonumber \\
\zeta_2 = & \ \frac{1}{\sigma_2 \sqrt{T}} \Bigg( \ln \{C_2(0)/F_2\} + (r - \frac{1}{2} \sigma_2^2) T \Bigg).   \nonumber 
\end{align}


\end{document}